\documentclass[twocolumn]{aastex631}
\pdfoutput=1
\usepackage{CJK}
\usepackage{relsize}
\usepackage{soul}

\usepackage[autostyle]{csquotes}
\usepackage{subfiles}
\usepackage{soul}
\usepackage{float}
\graphicspath{{./}{figures/}}
\usepackage{amsmath}
\usepackage{soul}
\usepackage{multirow}
\usepackage{hyperref}

\accepted{September 22, 2021}

\submitjournal{ApJ}

\graphicspath{{./}{figures/}}
\begin{document}
\begin{CJK*}{UTF8}{gbsn}
\shortauthors{Phillips et al.}
\author[0000-0001-5610-5328]{Caprice L. Phillips}
\affil{Department of Astronomy, The Ohio State University, 100 W 18th Ave, Columbus, OH 43210 USA}
\footnote{LSSTC DSFP Fellow}

\author[0000-0002-4361-8885]{Ji Wang (王吉)}
\affiliation{Department of Astronomy, The Ohio State University, 100 W 18th Ave, Columbus, OH 43210 USA}
\shorttitle{NH$_{3}$ Biosignature in Gas Dwarfs}

\author{Sarah Kendrew}
\affil{European Space Agency, Space Telescope Science Institute, 3700 San Martin Drive, Baltimore MD 21218, USA}

\author{Thomas P. Greene}
\affil{Space Science and Astrobiology Division, NASA Ames Research Center, MS 245-6, Moffett Field, CA 94035, USA}
\author{Renyu Hu}
\affil{Jet Propulsion Laboratory, California Institute of Technology, Pasadena, CA 91109, USA}
\affil{Division of Geological and Planetary Sciences, California Institute of Technology, Pasadena, CA 91125, USA}
\author{Jeff Valenti}
\affil{European Space Agency, Space Telescope Science Institute, 3700 San Martin Drive, Baltimore MD 21218, USA}
\author{Wendy {R.} Panero}
\affil{School of Earth Sciences, The Ohio State University, 125 South Oval Mall, Columbus OH, 43210, USA}
\author{Joe Schulze}
\affil{School of Earth Sciences, The Ohio State University, 125 South Oval Mall, Columbus OH, 43210, USA}

\title{Detecting Biosignatures in the Atmospheres of Gas Dwarf Planets with the James Webb Space Telescope}

\begin{abstract}

Exoplanets with radii between those of Earth and Neptune have stronger surface gravity than Earth, and can retain a sizable hydrogen-dominated atmosphere. In contrast to gas giant planets, we call these planets gas dwarf planets.  The James Webb Space Telescope (JWST) will offer unprecedented insight into these planets. Here, we investigate the detectability of ammonia (NH$_{3}$, a potential biosignature) in the atmospheres of seven temperate gas dwarf planets using various JWST instruments. We use \texttt{petitRadTRANS} and \texttt{PandExo} to model planet atmospheres and simulate JWST observations under different scenarios by varying cloud conditions, mean molecular weights (MMWs), and NH$_{3}$ mixing ratios.  A metric is defined to quantify detection significance and provide a ranked list for JWST observations in search of biosignatures in gas dwarf planets. It is very challenging to search for the 10.3--10.8 $\mu$m  NH$_{3}$ feature using eclipse spectroscopy with MIRI in the presence of photon and a systemic noise floor of 12.6 ppm for 10 eclipses. NIRISS, NIRSpec, and MIRI are feasible for transmission spectroscopy to detect NH$_{3}$ features from 1.5 $\mu$m to 6.1 $\mu$m under optimal conditions such as a clear atmosphere and low MMWs for a number of gas dwarf planets. We provide examples of retrieval analyses to further support the detection metric that we use. Our study shows that searching for potential biosignatures such as NH$_{3}$ is feasible with a reasonable investment of JWST time for gas dwarf planets given optimal atmospheric conditions.
\end{abstract}

\keywords{Exoplanet atmospheres, Exoplanets: Mini Neptunes -- Super Earths , Astrobiology - biosignatures}
\section{Introduction}
\label{sec:introduction}
The Kepler Space Mission \citep{borucki2010} has shown that super-Earths/mini-Neptunes are amongst the most abundant type of planet (\citealt{fressin2013}; \citealt{fulton2017}). However, their formation history, internal and atmospheric composition, and chemistry remain poorly understood due to their relatively small size and the presence of clouds~\citep{Madhusudhan2020,Benneke2019b}. 
After the $Kepler$ mission, The Transiting Exoplanet Survey Satellite ($TESS$; \citealt{ricker2015}), has provided more super-Earths/mini-Neptunes for characterization and future study of atmospheric composition (\citealt{Chouqar2020};  \citealt{FortenbachandDressing2020}). 

\par
As a successor to the Hubble Space Telescope (HST), the  James Webb Space Telescope (JWST) - with its larger collection area -  will allow for higher resolution and increased wavelength coverage to probe the atmospheric compositions of transiting super-Earths/mini-Neptunes.
\par
With a growing list of potentially habitable planets, the search for biosignatures is the next logical step in exoplanet studies. Biosignatures such as O$_{2}$ and CH$_{4}$ are familiar to Earth-like planets \citep{DesMarais2002}. However, there are a limited number of Earth-sized planets~\citep[e.g., TRAPPIST-1 d and e,][]{Gillon2017} that can be accessed by JWST. Moreover, the observation is very challenging in terms of the signals ($\sim10$ ppm) and thus the required telescope time~\citep{Yaeger2019, Suissa2020}.

Instead, we focus on gas dwarf planets \citep{Buchave2014}, which we define as super-Earths/mini-Neptunes that have hydrogen-dominated atmospheres with radii larger than 1.7 R$_{\oplus}$ and extend to 3.9 R$_{\oplus}$\footnote {Beyond 3.9 R$_{\oplus}$, these objects are considered ice or gas giants based on the host star metallicities \citep{Buchave2014}}. Beyond 1.7 R$_{\oplus}$, planets lie in the second bimodal distribution of the radius valley  and are more likely to have a gaseous envelope (\citealt{fulton2017};\citealt{VanEylen2018}). Gas dwarf planets are more amenable targets than Earth-like planets for transit observations because of larger radii. In addition, gas dwarf planets have larger atmosphere scale height because of the lower mean molecular weight (MMW) due to the H-dominate atmosphere, which further increases the transit signals.

\par
Because of the H-dominated atmosphere, gas dwarf planets have different atmospheric chemistry. The dominance of hydrogen creates a reducing chemistry, causing other elements to preferentially react with hydrogen to produce molecules such as water (H$_2$O), ammonia (NH$_3$), and methane (CH$_4$). This is contrast with the oxidizing chemistry that exists on the inhabited Earth. We therefore expect different biosignatures in gas dwarf planets. 

\cite{seagarbiomass} first proposed NH$_3$ as a biosignature gas in a H$_{2}$ and N$_{2}$ dominated atmosphere, nicknamed a \enquote{cold Haber World}, the reaction is as follows: $$3H_{2} + N_{2} \longrightarrow 2NH_{3}$$

NH$_3$ is a strong candidate as a biosignature for the following reasons~\citep{seagarbiomass}: first, the reaction that produces NH$_3$ from N$_2$ and H$_2$ is exothermic, i.e., releasing energy that can be harnessed by life to support metabolism,  second, this reaction requires high temperatures ($>$450 K) and high pressures ($>$10 bar) in abiotic environments, and therefore the existence of NH$_3$ at low temperatures and pressures in the upper atmosphere implies the existence of a reaction catalyst, potentially developed by life for metabolic processes, and
third NH$_3$ is easily destructible in photochemistry and volcanic environments. Therefore, any NH$_3$ has to be replenished by certain productive processes that potentially involve life.

While NH$_3$ is a promising biosignature for gas dwarf planets, the detection of NH$_3$ only provides a necessary condition for an inhabited ``cold Haber World" and sets the stage for follow-up observational work to confirm the detection and theoretical work to exhaust the chemical reaction network in order to ensure the production of NH$_3$ is only made possible by life. The extreme challenge in carrying out these efforts has been highlighted in the recent controversy on the detection of PH$_3$ in the atmosphere of Venus~\citep{Greaves2020, Greaves2020b, Snellen2020, Villanueva2020, Akins2021, Lincowski2021, Thompson2021}. 

Along the same cautionary note, gas dwarfs are generally not considered habitable unless there is a substantial ocean layer with a relative thin atmosphere  (\citealt{Madhusudhan2020}; \citealt{Scheucher2020}; \citealt{mousis2020irradiated};  \citealt{hu2021photochemical}; \citealt{nixon2021deep}). Such ocean/Hycean worlds can exist, e.g., K2-18~b  $\&$ TOI-270~d~(\citealt{Madhusudhan2020}; \citealt{MadhusudhanHyean2021}). However, we discuss gas dwarfs in this paper because there may be possibilities for floating microbial life~\citep{Seager2020}. Biosignature aside, the investigation of physical and chemical conditions of hydrogen-dominated atmospheres is in itself an exciting research subject. 

In this paper, we study the feasibility of using JWST to search for NH$_3$ in the atmospheres of gas dwarf planets. In Section \ref{sec:target_sample_selection} we describe the selection criteria for the targets in this study. We describe the model and simulations of transmission and emission spectra of targets in Section \ref{sec:simulating}. We discuss the JWST simulations  and introduce a detection metric for NH$_{3}$ in Section \ref{sec:simulating_JWST}, and discuss our results and various factors that affect the detection of ammonia in Section \ref{sec:main_results}. To validate the detection metric that we use in this work, we provide examples of atmospheric retrieval in \S \ref{sec:atmospheric_retrievals} to show that NH$_3$ and H$_2$O can be detected in optimal conditions. Discussions and conclusions are provided in Section \ref{sec:discussions} $\&$ \ref{sec:conclusions}.

\section{Sample Selection}
\label{sec:target_sample_selection}
We have compiled a list of  targets that would be optimal for observations following the launch of JWST. We use the NASA Exoplanet Archive (NEA)\footnote{\url{https://exoplanetarchive.ipac.caltech.edu}} and the following selection criteria: (1) planet radii between 1.7 and 3.4 R$_\oplus$; (2) equilibrium temperature (T$_{\rm{eq}}$) below 450 K; and (3) distance within 50 pc. 

The 1.7 R$_\oplus$ radius cut makes it more likely that our candidates have a gaseous envelope~\citep{Rogers2015}. Additionally, all but two planets in our sample live above the period-dependent radius gap \citep{VanEylen2018} further suggesting that these planets are indeed gas dwarfs/sub-Neptunes. The upper limit on radius reduces the likelihood that planets have a sufficiently high surface pressure to produce abiotic NH$_3$. We note, however, the surface pressure can vary by orders of magnitude depending on planet internal structure and atmospheric composition~\citep{Madhusudhan2020}. In Jupiter, the threshold pressure for chemical production of NH$_3$ is $\sim$1000 bar \citep{Prinn1981}. The upper limit of T$_{\rm{eq}}$ at 450 K is where liquid water can exist at 10-bar pressure~\citep{Chaplin2019}. Lastly, we select only nearby systems within 50 pc to ensure adequate flux from the star and planet.
\par
\begin{figure}[htp!]
    \centering
    \includegraphics[width=1.1\columnwidth]{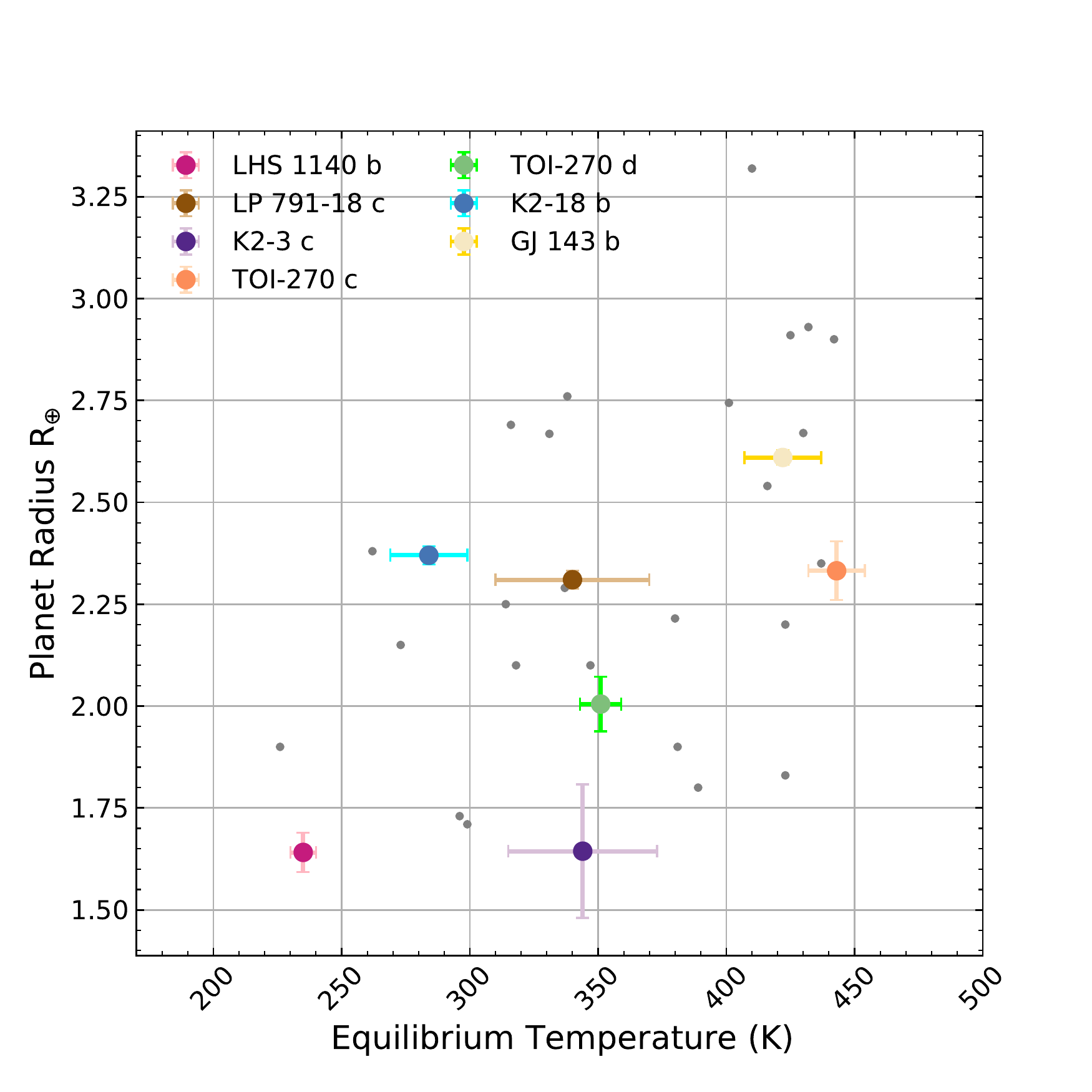}
    \caption{Planet radius (R$_{\oplus})$ vs equilibrium temperature (K) for LHS 1140 b, LP 791-18~c, K2-3~c, K2-18~b, TOI-270~c, TOI-270~d, and GJ 143 b. Lighter gray circles are those with distances $>$ 50 pc but still have radii between 1.7 and 3.4 Earth radii and T$_{eq}$ $<$ 450 K. Our seven colored targets are those that meet our full selection criteria in Section \ref{sec:target_sample_selection}.
    \label{fig:selected_targets}}
\end{figure}

Based on our selection criteria, we select seven targets for this work: LHS 1140 b, K2-3~c, TOI-270 c, TOI-270 d, K2-18~b, GJ 143 b, and LP 791-18 c (Figure \ref{fig:selected_targets}).
We summarize planetary and stellar parameters used in this study in Table \ref{tab:planetary_parameters}.

\begin{deluxetable*}{lccccccc}
\tablenum{1}
\tabletypesize{\scriptsize}

\tablecaption{Planetary and stellar Parameters of targets selected for this work }
\label{tab:planetary_parameters}
\tablehead{
\colhead{} & \colhead{LHS 1140 b}& \colhead{K2-3~c} & \colhead{TOI-270~c \tablenotemark{d}} & \colhead{TOI-270~d \tablenotemark{d}}& \colhead{K2-18~b }&\colhead{GJ 143 b \tablenotemark{e}} &\colhead{LP 791-18~c \tablenotemark{g}}}
\startdata
M$_{p}$ (M$_{\oplus}$)&6.96$\pm$0.89\tablenotemark{b}&2.14$^{+1.08}_{-1.04}$ (7) & 6.14$\pm$0.38 &4.78$\pm$0.46&8.92$^{+1.70}_{-1.60}$ (11)&22.7$^{+2.2}_{-1.9}$&5.96 (6) \tablenotemark{f}\\
R$_{p}$ (R$_{\oplus}$)&1.635$\pm$0.046 \tablenotemark{h} (2)&1.618$^{+0.212}_{-0.207}$\tablenotemark{i} (7)&2.332$\pm$0.072&2.005$\pm$0.007&2.3$\pm$0.22 (11)&2.61$^{+0.17}_{-0.16}$&2.31$\pm$0.25\\
Gravity (g) \tablenotemark{a} & 2.34 (6)&0.80 (6)&1.12 (6)&1.18 (6)&1.68 (6)&3.33 (6)& 1.11 (6)\\
T$_{eq}$ (K)&235$\pm$5 (1)& 344$\pm$29 (8) & 443$\pm$11&351$\pm$8&284$\pm$15 (11) & 422$^{+15}_{-14}$&370$\pm$30\\
Distance (pc)&14.98$\pm$0.01 (3)&44.07$\pm$0.10 (3)&\multicolumn{2}{c}{22.458$\pm$0.0059}&38.02$\pm$0.07 (3) & 16.32 $\pm$ 0.0071 (3) &26.4927$^{+0.0640}_{-0.0638}$\\
K-band$_{s}$ (mag)\tablenotemark{c}&8.821$\pm$0.024&8.561$\pm$0.023&\multicolumn{2}{c}{8.251$\pm$0.029}&8.99$\pm$0.02 & 5.524$\pm$0.031&10.644$\pm$0.023\\
T$_{s}$ (K)&3216$\pm$39 (1)&3896$\pm$189 (4)&\multicolumn{2}{c}{3506 $\pm$70}&3457$\pm$39 (12) &4640$\pm$100&2960$\pm$55\\
logg (dex)&5.03$\pm$0.02 (5)&4.734$\pm$0.062 (9)&\multicolumn{2}{c}{4.872$\pm$0.026}&4.856$\pm$0.062 (13) &4.613$^{+0.052}_{-0.061}$&5.115$\pm$0.094\\
t$_{14}$(hrs)&2.055$\pm$0.0046 (2)&3.38$\pm$0.12 (4) & 1.658$^{+0.015}_{-0.012}$(10)&2.14$\pm$0.018 (10)&2.663$^{+0.023}_{-0.027}$ (14)&3.20$^{+0.041}_{-0.038}$&1.208$^{+0.056}_{-0.046}$\\
Fe/H (dex)&-0.24$\pm$0.10 (1)&-0.32$\pm$0.13 (3)&\multicolumn{2}{c}{--0.20$\pm$0.12}&0.12$\pm$0.16 (11) & 0.003$\pm$0.06&-0.09$\pm$0.19\\
\enddata
\tablenotetext{a}{The gravity ($g_{p}$) and the subsequent pressure estimation for thermal emission spectra with \texttt{petitRADTRANS} is similar using the approximation from masses from NASA Exoplanet Archive}
\tablenotetext{b}{\cite{lillobox2020} find a RV measurement mass of 6.48 $\pm$ 0.46 M$_{\oplus}$}
\tablenotetext{c}{All values from Two Micron All Sky Survey (2MASS), unless otherwise noted
\citep{cutri2003}}
\tablenotetext{d}{All values from \cite{eylen2021} unless otherwise noted}
\tablenotetext{e}{All values from \cite{dragomir2019} unless otherwise noted}
\tablenotetext{f}{\cite{crossfield2019} estimate a mass of 7M$_{\oplus}$}
\tablenotetext{g}{All values from \cite{crossfield2019} unless otherwise noted}
\tablenotetext{h}{The radius of LHS 1140 b is very close to the 1.7 R$_{E}$ and it is one of the most well-characterized potential gas dwarf planets, so we include it in this study}
\tablenotetext{i}{We include K2-3~c because its radius uncertainty overlaps with the \cite{VanEylen2018} radius gap and our 1.7 R$_{\oplus}$ cutoff.}
\tablerefs{(1) \citealt{ment2019}, (2) \citealt{lillobox2020}, (3) \citealt{Gaia2}, (4) \citealt{crossfield2016}, (5) \citealt{stassun2019} (6) This work, (7) \citealt{kosiarek2019}, (8) \citealt{sinukoff2016}, (9) \citealt{Almenara2015}, (10) \citealt{gunther2019}, (11) \citealt{sarkis2018}, (12) \citealt{benneke2019}, (13) \citealt{crossfield2016}, (14) \citealt{benneke2017}}
\end{deluxetable*}
\vspace{-0.10in}
\section{Simulating Transmission and Emission Spectra}
\label{sec:simulating}
 We use the Python package, \texttt{petitRADTRANS}\footnote{\url{https://gitlab.com/mauricemolli/petitRADTRANS}} \citep{molliere2019}, which calculates emission and transmission spectra of exoplanets. The emission and transmission spectra are produced through the implementation of a radiative transfer code. The atmosphere is assumed to be plane-parallel and in local thermodynamic equilibrium. The open-source radiative transfer code allows for modification of pressure-temperature (P-T) profile (\S \ref{sec:pt}), atmospheric composition and MMW (\S \ref{sec:atmospheric_composition}), surface gravity and planet radius (\S \ref{sec:radiusandsurface}). The code can account for the effects of clouds, scattering, and collision induced absorption. 
 \par 
 There are two resolution modes available: low ($R=\frac{\lambda}{\Delta \lambda}$ = 1000) and high~($R=\frac{\lambda}{\Delta \lambda}$ = 10$^{6}$) modes. We utilize the low resolution mode, given that the Mid-Infrared Instrument (MIRI) LRS and Near-Infrared Spectrograph (NIRSpec), and NIRISS (SOSS) modes have resolutions of $\sim$ 100, 100 - 2700, and 700  respectively. In this study, we test a variety of cases for our sample: (1) no clouds, high-MMW emission spectra, (2) no clouds, low-MMW emission spectra, (3) no clouds, high-MMW transmission spectra, (4) no clouds, low-MMW transmission spectra,  and (5) cloud deck at 0.01 bar, 0.1 bar, and 1 bar, low-MMW transmission spectra.
\subsection{P-T Profiles}
\label{sec:pt}
For emission spectroscopy the P-T profiles for our targets are adjusted and based on the P-T profile of Earth. According to \citealt{Seager2013a} - who approximates a P-T profile for a \enquote{cold Haber World } with a T$_{eq}$ =  290 K - the precise temperature pressure structure of the atmosphere is less important than photochemistry for a first order description of biosignatures in H$_{2}$ rich atmospheres.  We utilize public atmospheric data available from Public Domain Aeronautical Software (PDAS) to produce a pressure temperature profile for Earth\footnote{\url{http://www.pdas.com/atmosdownload.html}}, and shift the profile based on the planet surface pressure so that 1 bar atmospheres matches the equilibrium temperature of the targets.

We use an isothermal P-T profile for the transmission spectroscopy. Unlike emission spectroscopy, an isothermal P-T profile will not produce a featureless transmission spectrum.  However, isothermal profile may be overly simplified and can introduce bias in retrieval analysis \citep{Rochetto2016}. 

\subsection{Atmospheric Composition}
\label{sec:atmospheric_composition}
We calculate the volume mixing ratio (VMR) using values for a \enquote{cold Haber World}~\citep{seagarbiomass} and the VMRs are reported in Table \ref{tab:superearth} $\&$ Table \ref{tab:superearth_low}. We assume the same chemistry of a \enquote{cold Haber World} to focus on the effects of temperature on the magnitude of transmission spectrum. The VMR  is calculated using the following equation:   
\begin{equation}
VMR = \cfrac{n_i}{\Sigma n_i},
\end{equation}
where $n_i$ is the mixing ratio from~\citep{seagarbiomass} for a given species at 1-bar pressure. 
 The species with opacity information in \texttt{petitRADTRANS} are: H$_{2}$O, CO$_{2}$, CH$_{4}$, H$_{2}$, CO, HCN, OH, and NH$_{3}$. In addition, we include H$_{2}$ and He for collision-induced absorption and N$_{2}$ as a filler gas. 
\par 

The VMRs from \cite{seagarbiomass}  are summed and  normalized by dividing by the summation  so the total VMR adds to 1.0.
\par
Since \texttt{petitRADTRANS} takes in mass mixing ratio (MMR), we calculate MMR as follows:
\begin{equation}
    {\rm{MMR}}_{i} = \cfrac{\mu_{i}}{\mu}{\rm{VMR}}_{i}
\end{equation}
where $\mu_{i}$ is the mass of the species in atomic unit, $\mu$ is the MMW for the atmosphere in atomic unit, and ${\rm{VMR}}_{i}$ is the volume mixing ratio. The MMW is calculated as:
\begin{equation}
    \mu =  \sum{\mu_{i}}\cdot{\rm{VMR}}_{i}.
\end{equation}
\begin{deluxetable}{ccc}[htp!]
\tablenum{2}
\tablecaption{Species used for \texttt{petitRADTRANS} to generate synthetic spectra with 25$\%$ H$_{2}$ and 75$\%$ N$_{2}$ \label{tab:superearth}}
\tablehead{
\colhead{Species} & \colhead{VMR} & \colhead{MMR}}
\startdata
H$_{2}$O &9.23e-07&8.24e-07\\
CO$_{2}$&2.92e-09&6.37e-09\\
CH$_{4}$&2.92e-08&2.31e-08\\
H$_{2}$&2.30e-01&2.29e-02\\
CO&9.23e-10&1.28e-09\\
OH&9.23e-16&7.78e-16\\
HCN&9.23e-10&1.23e-09\\
NH$_{3}$&3.69e-06&3.11e-06\\
He&7.69e-02&1.52e-02\\
N$_{2}$\tablenotemark{a}&6.92e-01&9.61e-01
\enddata
\tablenotetext{a}{N$_{2}$ has no rotational-vibrational transitions, so there are no spectral signatures visible at infrared wavelengths, so this feature is not available in \texttt{petitRADTRANS} but is used to determine mean molecular weight of atmosphere}
\end{deluxetable}

\begin{deluxetable}{ccc}[htp!]
\tablenum{3}
\tablecaption{Species used for \texttt{petitRADTRANS} to generate synthetic spectra with 90$\%$ H$_{2}$  and 10$\%$ N$_{2}$ \tablenotemark{a} atmosphere  \label{tab:superearth_low}}
\tablehead{
\colhead{Species} & \colhead{VMR} & \colhead{MMR}}
\startdata
H$_{2}$O &9.17e-07&3.62e-06\\
CO$_{2}$&2.90e-09&2.81e-08\\
CH$_{4}$&2.90e-08&1.02e-07\\
H$_{2}$&8.25e-01&3.62e-01\\
CO&9.17e-10&5.64e-09\\
OH&9.17e-16&3.42e-15\\
HCN&9.17e-10&5.44e-09\\
NH$_{3}$&3.66e-06&1.37e-05\\
N$_{2}$\tablenotemark{b}&9.17e-02&5.64e-01\\
He&8.25e-02&7.25e-02
\enddata
\tablenotetext{a}{We simply adopt the mixing ratios for gases other than N$_{2}$ or H$_{2}$, assuming no major change in the chemistry}
\tablenotetext{b}{N$_{2}$ has no rotational-vibrational transitions, so there are no spectral signatures visible at infrared wavelengths, so this feature is not available in \texttt{petitRADTRANS} but is used to determine mean molecular weight of atmosphere}
\end{deluxetable}
\vspace{-0.10in}
\subsection{Radius and Surface gravity}
\label{sec:radiusandsurface}
Radius and surface gravity are also the input for \texttt{petitRADTRANS}. The radii are taken from the NASA Exoplanet Archive (Table \ref{tab:planetary_parameters}). To calculate surface gravity, we also get masses from NEA, except for the mass of LP 791-18 c for which we use a mass-radius scaling relation~\citep{crossfield2019}. 

\subsection{Emission Spectroscopy}
\begin{figure}[htp!]
    \centering
    \includegraphics[width=1.1\columnwidth]{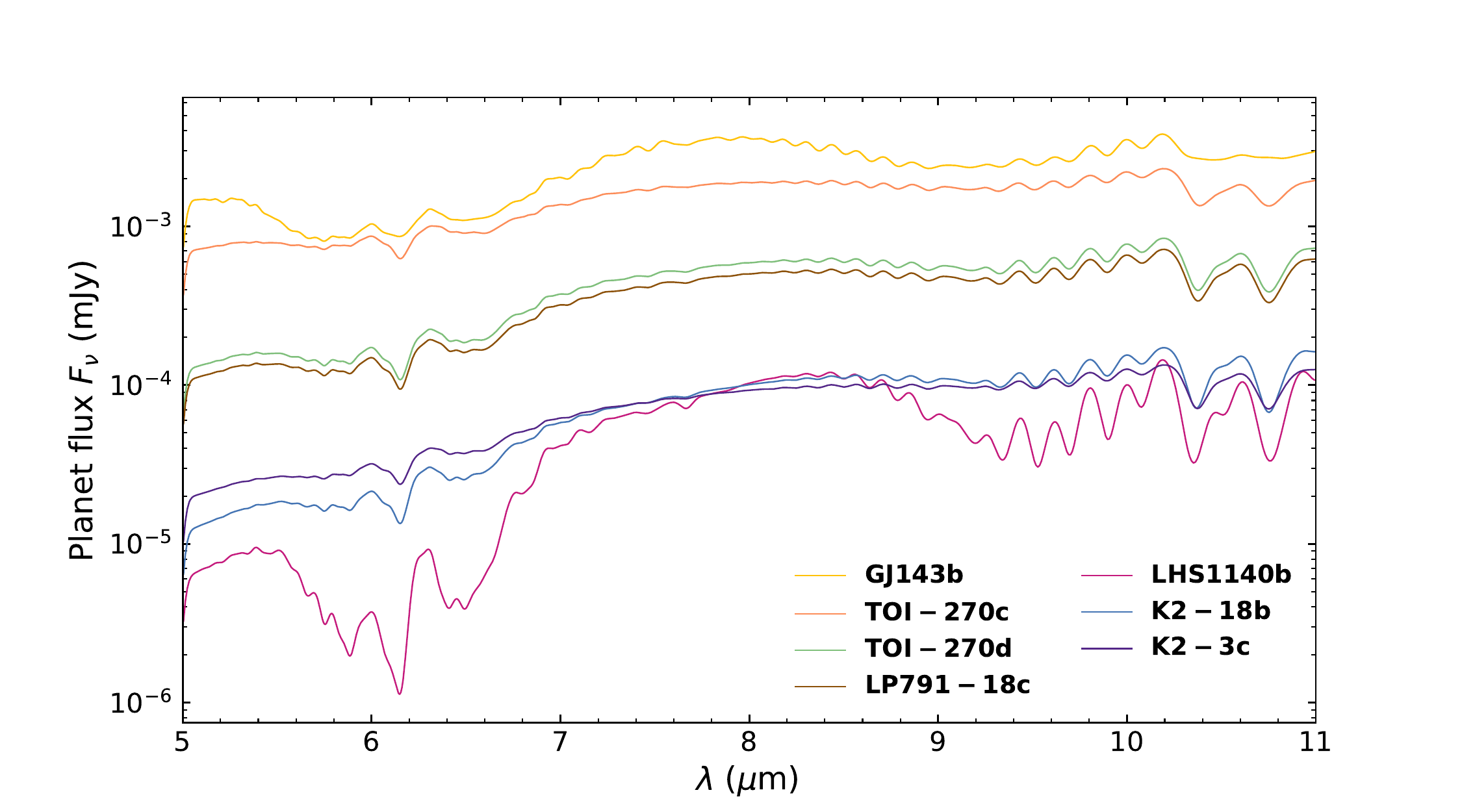}
     \caption{Synthetic  emission spectra of select targets generated from \texttt{petitRADTRANS}: GJ 143 b, TOI-270 c, TOI-270 d, LHS 1140 b, K2-18 b, LP 791-18 c,  and K2-3~c. The spectra have been convolved with a 1D Gaussian kernel down to R $\sim$ 100 from R$\sim$1000 to match the resolution of MIRI LRS.}
     \label{fig:emission_all_targets}
\end{figure}
With the inputs that are described in previous sections, we model the emission spectra ($f_{p}$) as shown in Figure \ref{fig:emission_all_targets}.
 The output flux unit for \texttt{petitRADTRANS} in mJy is converted to $f_{p}/f_{\star}$, which is the input for \texttt{PandExo}~\citep{batalha2017} to generate simulated JWST data.  

\begin{figure}[htp!]
    \centering
   \includegraphics[width=1.1\columnwidth]{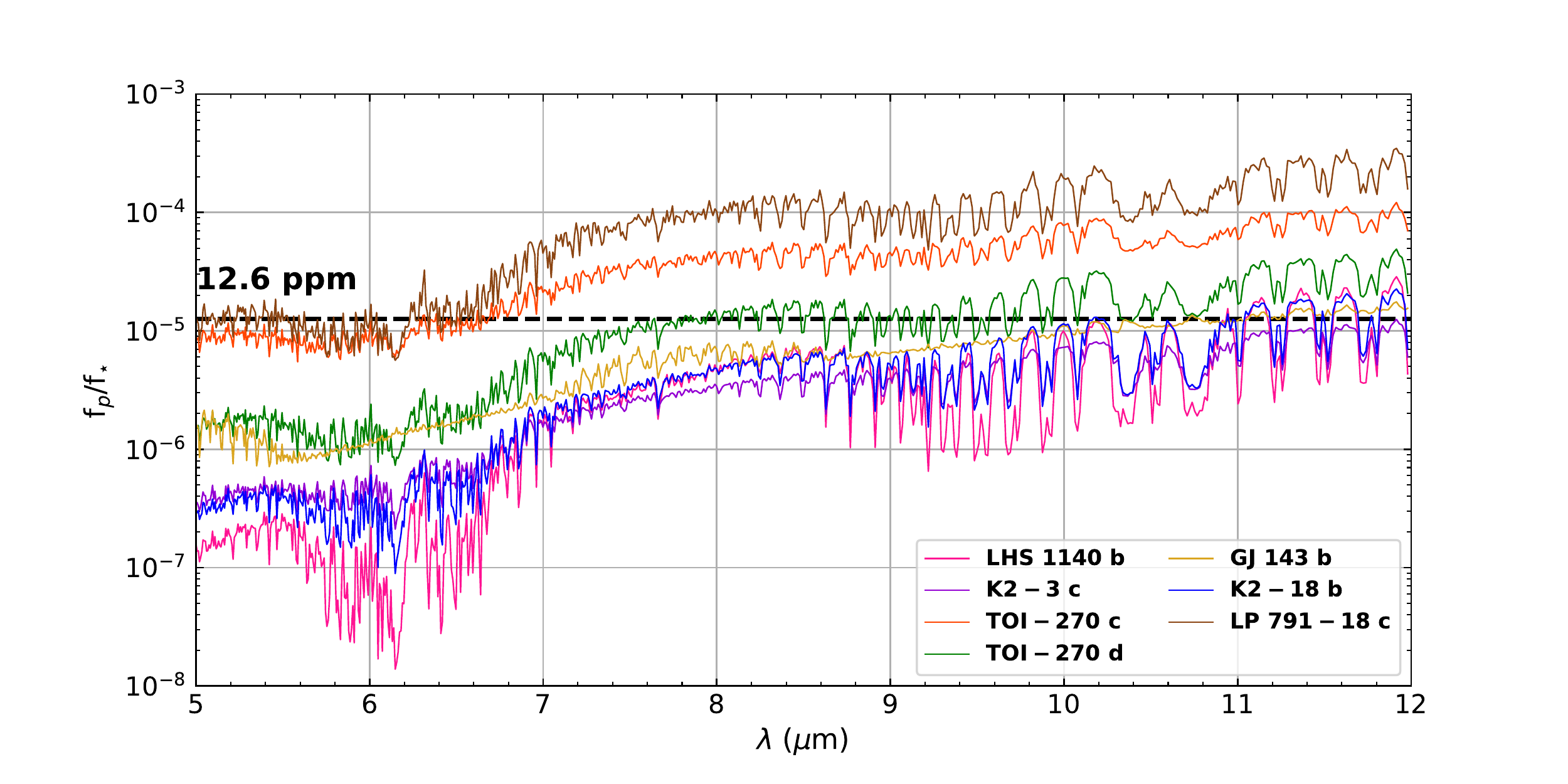}
    \caption{Simulated flux ratio ($f_{p}/f_{\star}$) comparisons for LHS 1140 b, K2-3~c, TOI-270 c, TOI-270 d, GJ 143 b, K2-18 b, and LP 791-18 c for a 90$\%$H$_{2}$/10$\%$N$_{2}$, no clouds atmosphere. The dashed line represents the systemic noise limit for MIRI LRS for 10 transits, where the noise level shown (12.6 ppm) follows as 40/$\sqrt{N_{obs}}$, where $N_{obs}$ = 10.}
   \label{fig:fpfstar}
\end{figure}

To calculate stellar flux ($f_{\star}$), we use the PHOENIX model grids \citep{Husser2013}. We re-sample the wavelength grid of the synthetic planet spectrum from \texttt{petitRADTRANS} onto the wavelength grid of the PHOENIX model to determine $f_{p}/f_{\star}$. Examples of planet-star contrast spectra are shown in Figure \ref{fig:fpfstar}.

\subsection{Transmission Spectroscopy}
\begin{figure}[htp!]
    \centering
    \includegraphics[width=1.1\columnwidth]{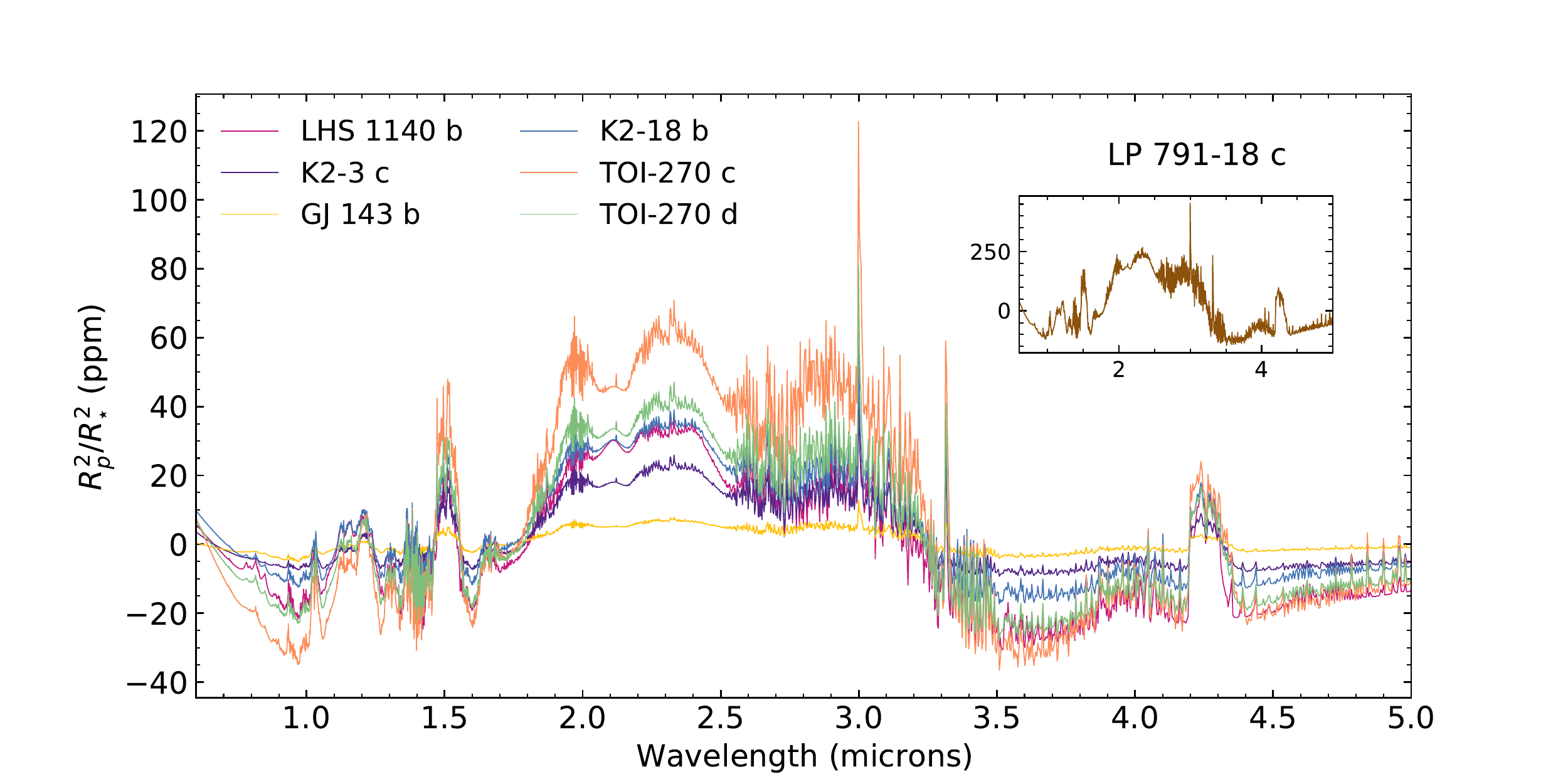}
    \caption{Transmission signal (minus offset) for a 90$\%$H$_{2}$/10$\%$N$_{2}$, no clouds atmosphere for targets in our sample. LP 791-18~c has the highest transmission signal and is shown in a separate plot to the side.}
    \label{fig:transmission_all_targets}
\end{figure}

Some examples of the modeled transmission spectra are shown in  Figure \ref{fig:transmission_all_targets}. The reference pressure for all targets are set to $P_{0}$ = 1.0 bar and the planet radius and surface gravity values (Table \ref{tab:planetary_parameters}) are given at $P_{0}$ in \texttt{petitRADTRANS}. 
\section{Simulating JWST Observations}
\label{sec:simulating_JWST}
\subsection{Considered Instruments}
 We consider NIRSpec, MIRI, and NIRISS instruments for this work\footnote{All assumptions on the performance of these instruments are based on pre-launch, ground-test data and models}. A summary of the instrument specifications is provided in Table \ref{tab:instrument_summary}.  These instruments allow for a range of wavelengths that cover major NH$_{3}$ features near  1.0--1.5, 2.0, 2.3, 3.0, 5.5--6.5, and 10.3--10.8 $\mu$m, and spectral features from other molecular species such as CO, CO$_{2}$, HCN, and CH$_{4}$. NIRCam has lower throughput than NIRSpec, MIRI, and NIRISS so we omit this instrument from this study.
\begin{deluxetable}{ccccc}[htp!]
\tablenum{4}
\tablecaption{Summary of instruments $\&$ modes }
\label{tab:instrument_summary}
\tablehead{\colhead{Instrument} & \colhead{Mode} & \colhead{Coverage} & \colhead{Resolution} & \colhead{Throughput \tablenotemark{b}}}
\startdata
NIRSpec&G235M\tablenotemark{a}&1.7--3.0 $\mu$m&1000&0.5\\
NIRSpec&G395M&2.9--5.0 $\mu$m&1000&0.6\\
NIRSpec&PRISM$\backslash$CLEAR&0.6--5.3 $\mu$m&100&0.5\\
NIRISS&SOSS&0.6--2.8 $\mu$m&700&0.3\\
MIRI&LRS$\backslash$slitless&5--12 $\mu$m&100&0.3
\enddata
\tablenotetext{a}{NIRSpec$\backslash$G235H has a higher resolution (R $\sim$ 2700) and comparable sensitivity but has a gap in the wavelength coverage so it is omitted from this study.}
\tablenotetext{b}{Estimated peak throughput values from JWST ETC version 1.6}
\end{deluxetable}

\begin{deluxetable}{cccc}[htp!]
\tablenum{5}
\tablecaption{Saturation limits for our targets \label{tab:saturated_limit}}
\tablehead{
\colhead{Target} & \colhead{Instrument} & \colhead{Mode} & \colhead{Saturation}}
\startdata
\multirow{4}{*}{LHS 1140 b}&NIRSpec&G235M&N\\
&NIRSpec&G395M&N\\
&NIRSpec&PRISM/CLEAR&Y\\
& MIRI&LRS/slitless&N\\
& NIRISS&SOSS&N\\
\hline
\multirow{4}{*}{LP 791-18~c}&NIRSpec&G235M&N\\
&NIRSpec&G395M&N\\
&NIRSpec&PRISM/CLEAR&N\\
& MIRI&LRS/slitless&N\\
& NIRISS&SOSS&N\\
\hline
\multirow{4}{*}{K2-3~c}&NIRSpec&G235M&N\\
&NIRSpec&G395M&N\\
&NIRSpec&PRISM/CLEAR&Y\\
& MIRI&LRS/slitless&N\\
& NIRISS&SOSS&N\\
\hline
\multirow{4}{*}{K2-18~b}&NIRSpec&G235M&N\\
&NIRSpec&G395M&N\\
&NIRSpec&PRISM/CLEAR&Y\\
& MIRI&LRS/slitless&N\\
& NIRISS&SOSS&N\\
\hline
\multirow{4}{*}{TOI-270~c}&NIRSpec&G235M&N\\
&NIRSpec&G395M&N\\
&NIRSpec&PRISM/CLEAR&Y\\
& MIRI&LRS/slitless&N\\
& NIRISS&SOSS&N\\
\hline
\multirow{4}{*}{TOI-270~d}&NIRSpec&G235M&N\\
&NIRSpec&G395M&N\\
&NIRSpec&PRISM/CLEAR&Y\\
& MIRI&LRS/slitless&N\\
& NIRISS&SOSS&N\\
\hline
\multirow{4}{*}{GJ 143~b}&NIRSpec&G235M&Y\\
&NIRSpec&G395M&Y\\
&NIRSpec&PRISM/CLEAR&Y\\
& MIRI&LRS/slitless&N\\
& NIRISS&SOSS&Y\\
\enddata
\end{deluxetable}
\vspace{-0.8in}
 \subsection{Eclipse Spectroscopy with MIRI}
 \label{sec:eclipse_spectroscopy}
 Ammonia has major absorption features at 10.3--10.8 $\mu$m, so we test the capabilities of MIRI LRS \citep{Kendrew2015} to detect this feature. We exclude the use of MIRI MRS because although this mode has a higher resolution, it has a lower throughput than MIRI LRS \citep{Glasse2015}. We note that MIRI MRS has been considered in other non-transiting exoplanet studies~\citep{Snellen2017}. The use of MIRI MRS for transiting exoplanets is still possible; capabilities of this mode will be further investigated in Cycle 1~(\citealt{TimeSeriesMRSKendrew}; \citealt{Deming2021JWST}).
 
 We use \texttt{PandExo} \citep{batalha2017} to simulate  observations for our targets.  We assume a stack of 10 eclipse observations, a 80$\%$ detector saturation level, and a floor noise of 40 ppm. The floor noise is set following \citet{Chouqar2020}, adopting a value of 40 ppm between the 30 ppm reported in~\citet{beichman2014} and 50 ppm reported in~\citet{Greene2016}. The noise floor in \texttt{PandExo} is only for 1 transit, and goes as $1/\sqrt{N}$ for stacking $N$ eclipses.
 \par
To further set up a run in PandExo, we input our simulated $f_{p}/f_{\star}$ spectra and PandExo uses the incorporated PHOENIX model library~\citep{Husser2013} as the stellar input for the simulated observations  based on the effective temperature, surface gravity, $K$-band magnitude, and metallicity [Fe/H] for each target host star.

\subsection{Detection Metric for Transmission Spectroscopy}
We define the significance of spectral feature detection using a signal to noise ratio (S/N): 
\begin{equation}
\label{eq:S/N}
    S/N = \cfrac{(R_{p}/R_{\star})^{2}-\overline{{(R_{p}/R_{\star})^{2}}}}{\sigma_{(R_{p}/R_{\star})^{2}}},
\end{equation}
\noindent where $(R_{p}/R_{\star})^{2}$ is the transmission signal from \texttt{petitRADTRANS},  $\overline{(R_{p}/R_{\star})^{2}}$ is median of the transmission signal from \texttt{petitRADTRANS}\footnote{We utilize wavelength ranges outside 1.0--3.5 $\mu$m to calculate the median as the range below 1.0  and above 3.5 do not include any major NH$_{3}$ absorption features.} and $\sigma_{(R_{p}/R_{\star})^{2}}$ is the uncertainty. 
\par
\begin{figure}[htp!]
    \centering
    \vspace{-0.25in}
    \includegraphics[width=1.1\columnwidth]{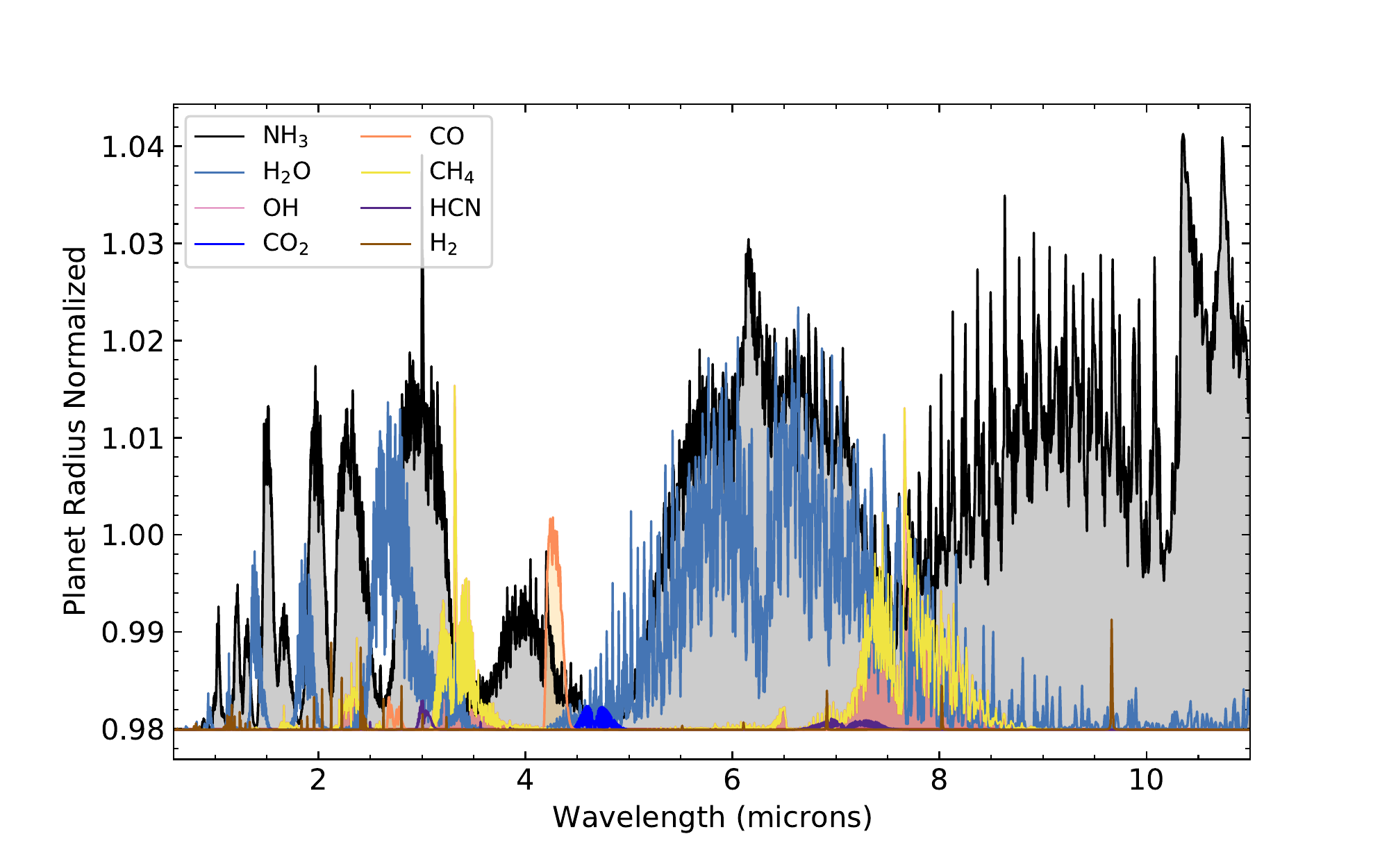}
    \caption{Near and mid-infrared wavelength ranges of major features of NH$_{3}$, H$_{2}$O, OH, CO, CO$_{2}$, CH$_{4}$, HCN and H$_{2}$.} 
    \label{fig:over_lapping}
\end{figure}

Following \cite{wunderlich2020} and \cite{Chouqar2020} we note ammonia features in the near and mid-infrared have other spectral features that overlap and can obscure these feature (Figure \ref{fig:over_lapping}). For example, the 2.0$\mu$m  NH$_{3}$  feature overlaps with H$_{2}$0 and H$_{2}$-H$_{2}$ features. In the mid-infrared the 10.3--10.8 $\mu$m NH$_{3}$ feature overlaps with H$_{2}$-H$_{2}$~\citep{wunderlich2020}. 

Similar to \cite{Chouqar2020}, we neglect the complications from these overlapping features of H$_2$O, NH$_3$ and other species. Our detection metric focuses on the S/N for detecting any spectral features, whether or not they are overlapped. The metric will help in determining the number of transits to significantly detect potential biosignatures for given atmospheric compositions. In addition, we show in \S \ref{sec:atmospheric_retrievals} that H$_2$O and NH$_3$ can be independently detected and constrained in atmospheric retrieval analyses despite overlapping spectral features.  
\vspace{-0.10in}
\subsection{Transmission Spectroscopy with NIRSpec and NIRISS}
To simulate JWST transmission spectra, we utilize the Near InfrarRed Spectrograph (NIRSpec)  instrument with the G235M and G395M modes which covers the wavelength ranges of 1.7--3.0 $\mu$m and 2.9--5.0 $\mu$m respectively. NIRSpec/G395M mode has an expected floor noise of 25 ppm \citep{Kreidberg2014}. We use a fixed number of 10 transits for our simulations with
\texttt{PandExo} as with the MIRI LRS simulation.

We assume an optimistic noise floor levelfor NIRISS (SOSS) and the NIRSpec modes. \cite{Greene2016} notes that \enquote{The best HST WFC3 G141 observations of transiting systems to date have noise of the order of 30 ppm \citep{Kreidberg2014}...}. The noise floor in \texttt{PandExo} is only for 1 transit, and goes as $1/\sqrt{N}$ for stacking $N$ transits.
\par
We also consider the use of the Near Infrared Imager and Slitless Spectrograph (NIRISS) instrument in the Single Object Slitless Spectroscopy (SOSS) mode which covers the 0.8--2.8 $\mu$m range. This compliments the wavelength coverage for the NIRSpec/G235M and NIRSpec/G395M modes. An optimistic noise floor of 20 ppm is adopted for NIRISS (SOSS) ~(\citealt{Greene2016}; \citealt{FortenbachandDressing2020}). 
\par
\texttt{Pandexo} simulations for these instruments for LHS 1140 b, LP 791-18~c, TOI-270~d, LHS 1140 b, K2-18~b, K2-3~c, and GJ 143 b are provided in the Appendix.

\subsection{Transmission Spectroscopy with MIRI LRS}
We also simulate whether the 6.1 and 10.3--10.8 micron feature of ammonia is detectable with transmission spectroscopy with MIRI LRS, using our detection metric. We assume the same setup for \texttt{PandExo} as with \S \ref{sec:eclipse_spectroscopy}. 
\section{Main Results}
\label{sec:main_results}

\subsection{Eclipse Spectroscopy with MIRI}
\label{sec:miri_eclipse}

\begin{figure}[htp!]
    \centering
    \vspace{-0.10in}
  \includegraphics[width=1.1\columnwidth]{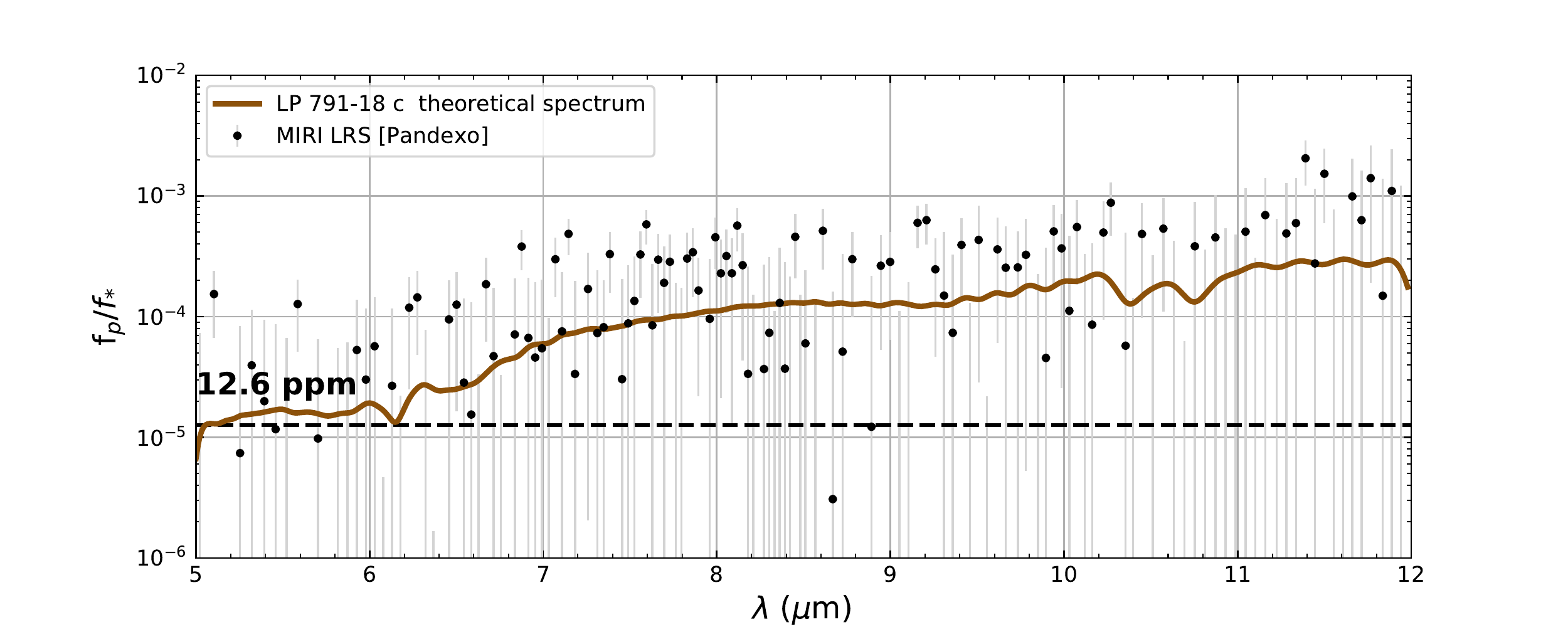}
  \includegraphics[width=1.1\columnwidth]{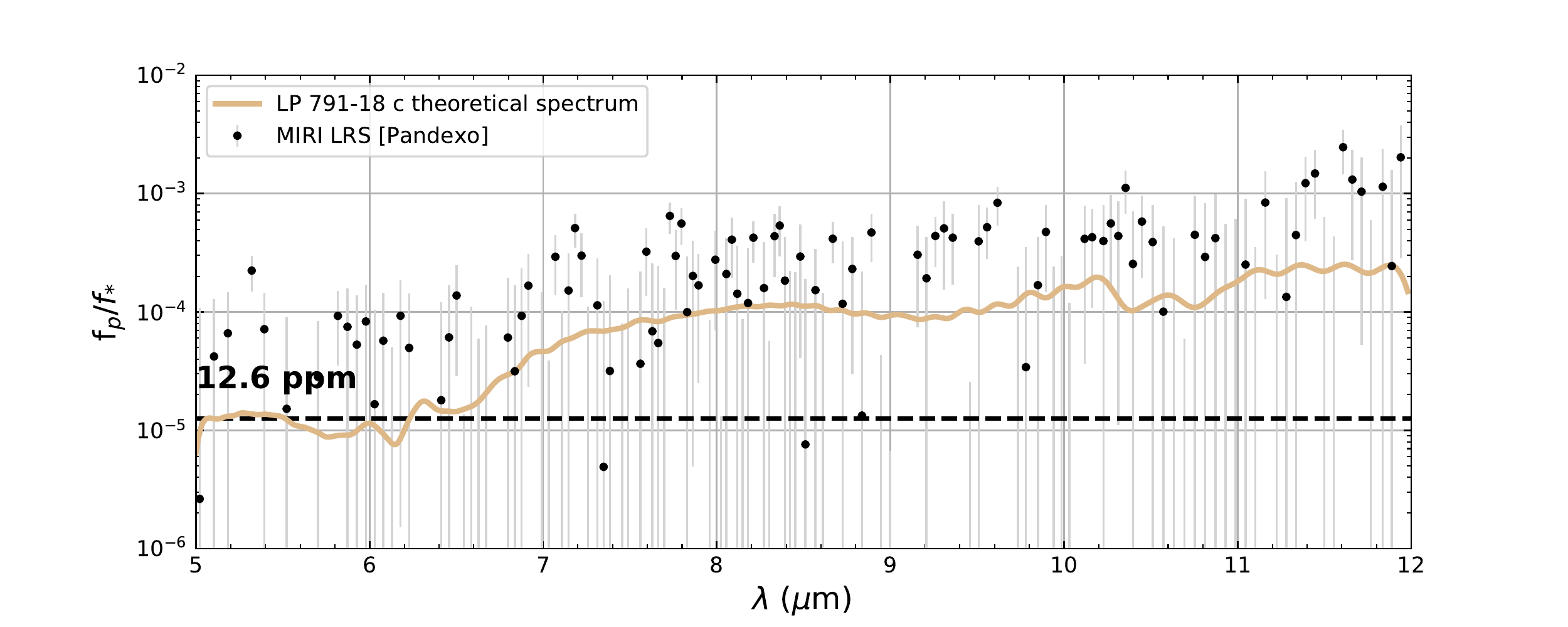}
    \caption{Simulated MIRI LRS emission spectra for 10 transits of LP 791-18 c
    The spectra from \texttt{petitRADTRANS} (R$\sim$1000) has been  smoothed down to match the resolution of MIRI LRS (R$\sim$100). he dashed line represents the systemic noise limit for MIRI LRS for 10 transits, where the noise level shown (12.6 ppm) follows as 40/$\sqrt{N_{obs}}$, where $N_{obs}$ = 10. \textit{Top}: High MMW (25$\%$ H$_{2}$) atmosphere with no clouds. \textit{Bottom:} Low MMW (90$\%$ H$_{2}$) atmosphere with no clouds.}
    \label{fig:emission_spectra_lp}
\end{figure}

The majority of our targets are not feasible for eclipse spectroscopy with MIRI LRS (Figure \ref{fig:fpfstar}). 
This is consistent with other studies that achieving the necessary flux contrast for significant detection of molecular features has proved to be very difficult with MIRI (\citealt{batalha2018}; \citealt{Chouqar2020}). 
LP 791-18 c has the most promising flux contrast ratio but the emission signal is mostly overwhelmed by the photon noise, assuming 10 transits and $\sim$ 40 hours of observing time (Figure \ref{fig:emission_spectra_lp}). Similarly TOI-270~c and TOI-270~d are photon limited. LP 791-18~c, TOI-270~c, and TOI-270~d have a S/N of 0.3$\sigma$, 0.2$\sigma$, and 0.1$\sigma$ for the 10.3--10.8 NH$_{3}$ feature. K2-3~c, LHS 1140~b, K2-18~b, and GJ 143~b are limited by systemic noise as shown in Figure. \ref{fig:fpfstar}.
\par

\subsection{Transmission Spectroscopy with NIRISS, NIRSpec, and MIRI}
\label{sec:all_trans}

Based on our transmission spectroscopy  detection metric (Equation \ref{eq:S/N}), we find that the NIRSpec/G235M, NIRSpec/G395M and NIRISS/SOSS modes are best suited to detect ammonia for our targets, with one exception --- LP 791-18 c for which NIRSpec PRISM/CLEAR is optimal because this target does not saturate the detector~(Table \ref{tab:saturated_limit}). 

\par
We compile a ranked list of targets based on transmission spectroscopy simulations for the NIRSpec/G235M, NIRSpec/G395M, NIRISS/SOSS, and MIRI LRS modes with 10 transits. To compute the rank list (Table \ref{tab:ranking}),  we use six major absorption features of NH$_{3}$ (1.5, 2.0, 2.3,  3.0, 6.1, and 10.3--10.8 $\mu$m). 
\par
For the 1.5, 2.0, 2.3, 3.0, 6.1, and 10.3--10.8 $\mu$m NH$_{3}$ features we find the approximate central wavelength and use three data points [one centered on the approximate central wavelength and two adjacent data points and compute the S/N based on the average of the three data points. 

\par
For the final S/N determination, we square each S/N of the NH$_{3}$ feature, take the summation, and take the square root of the summation (Equation \ref{eq:rank_signal}) to determine the ranking
\par

\begin{equation}
    \label{eq:rank_signal}
    <S/N> = \sqrt{\mathlarger{\sum_{i}}
    S/N_{i}^{2}}
\end{equation}
\noindent where $i$ indicates NH$_{3}$ $\lambda_{i}$ features at 1.5, 2.0, 2.3, 3.0, 6.1 and 10.3--10.8 $\mu$m.
\par
TOI-270~c is best suited for atmospheric studies with JWST given the S/N of detection features for NH$_{3}$. On the other hand, GJ 143 b is the least favorable target given that it  saturates the NIRSpec/G235M, NIRSpec/G395M, and NIRISS/SOSS modes due to the brightness of the host star (Table \ref{tab:saturated_limit}).
\par
\begin{figure}[htp!]
    \centering
    \includegraphics[width=1.1\columnwidth]{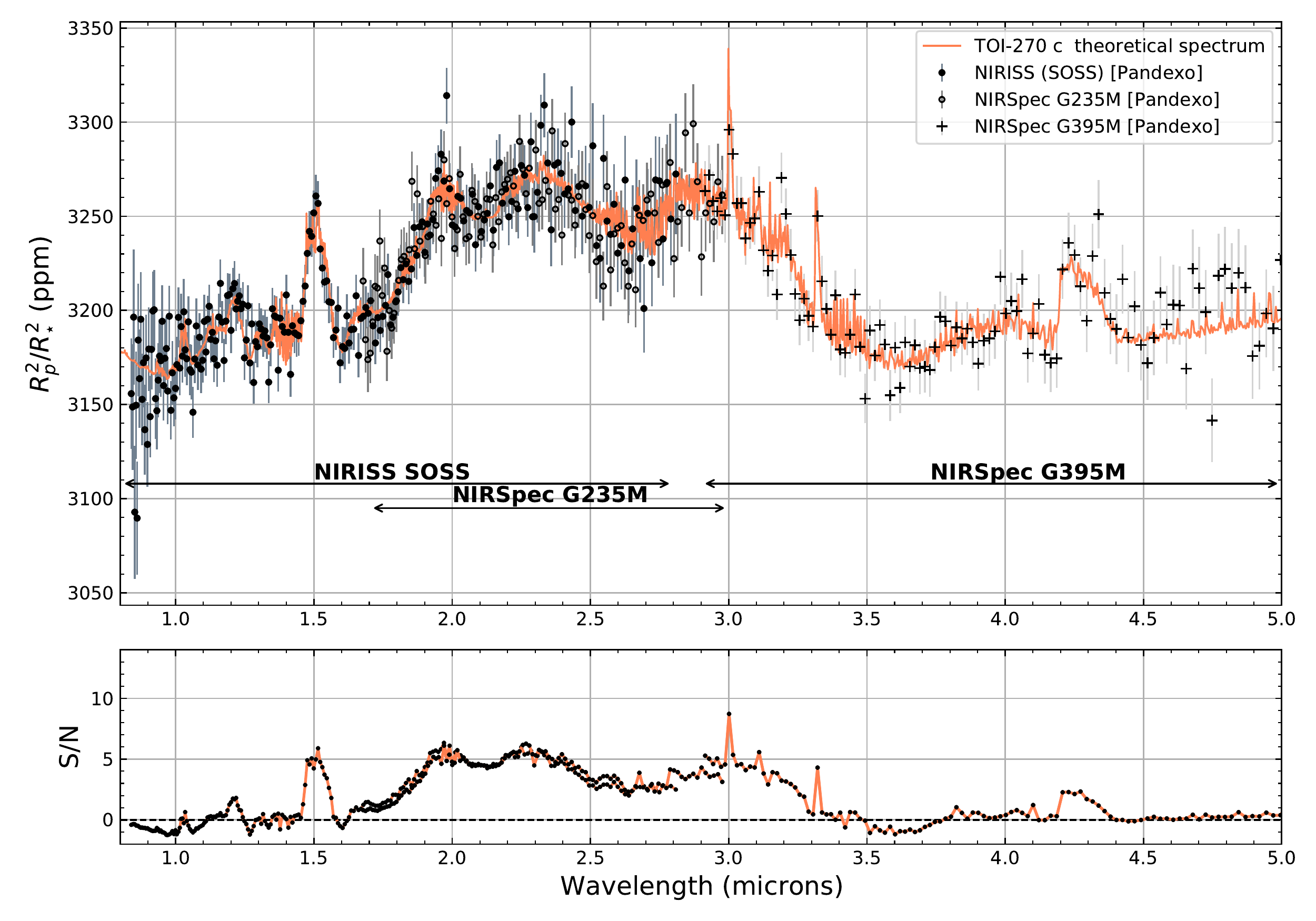}
    \caption{ Simulated NIRSpec/G235M, NIRSpec/G395M, and NIRISS (SOSS) transmission spectrum of TOI-270~c with 10 transits  and corresponding S/N for low MMW (MMW$\sim$ 4.55) (Table \ref{tab:superearth_low}).}
    \label{fig:lowmmw_tranmission}
\end{figure}

\begin{figure}[htp!]
    \centering
    \includegraphics[width=1.1\columnwidth]{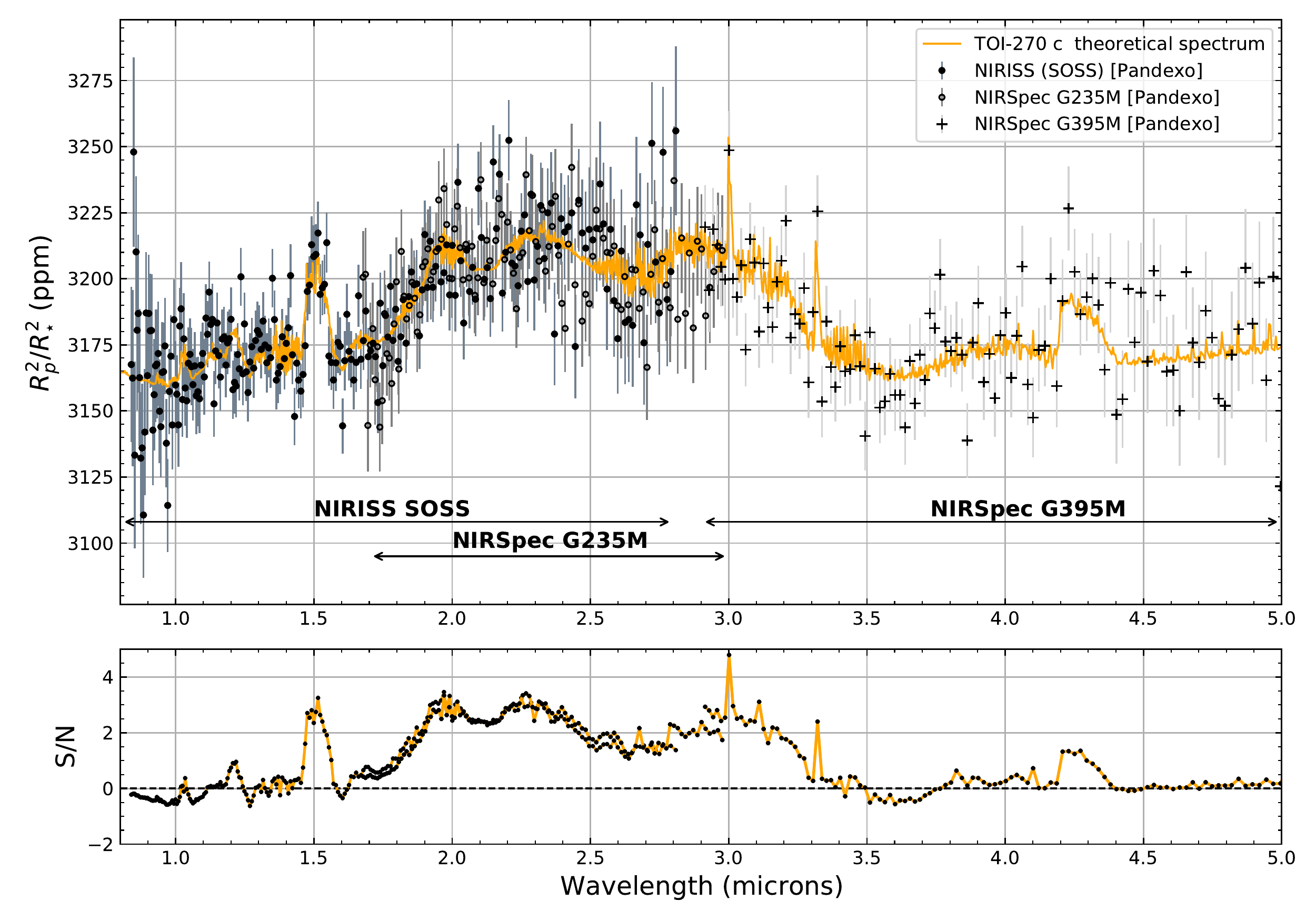}
    \caption{Simulated NIRSpec/G235M, NIRSpec/G395M, and NIRISS (SOSS) transmission spectrum of TOI-270~c with 10 transits and corresponding S/N for  the high MMW case MMW $\sim$ 20.1)(Table \ref{tab:superearth})}
    \label{fig:highMMW}
\end{figure}

\begin{figure}
    \centering
    \includegraphics[width=1.1\columnwidth]{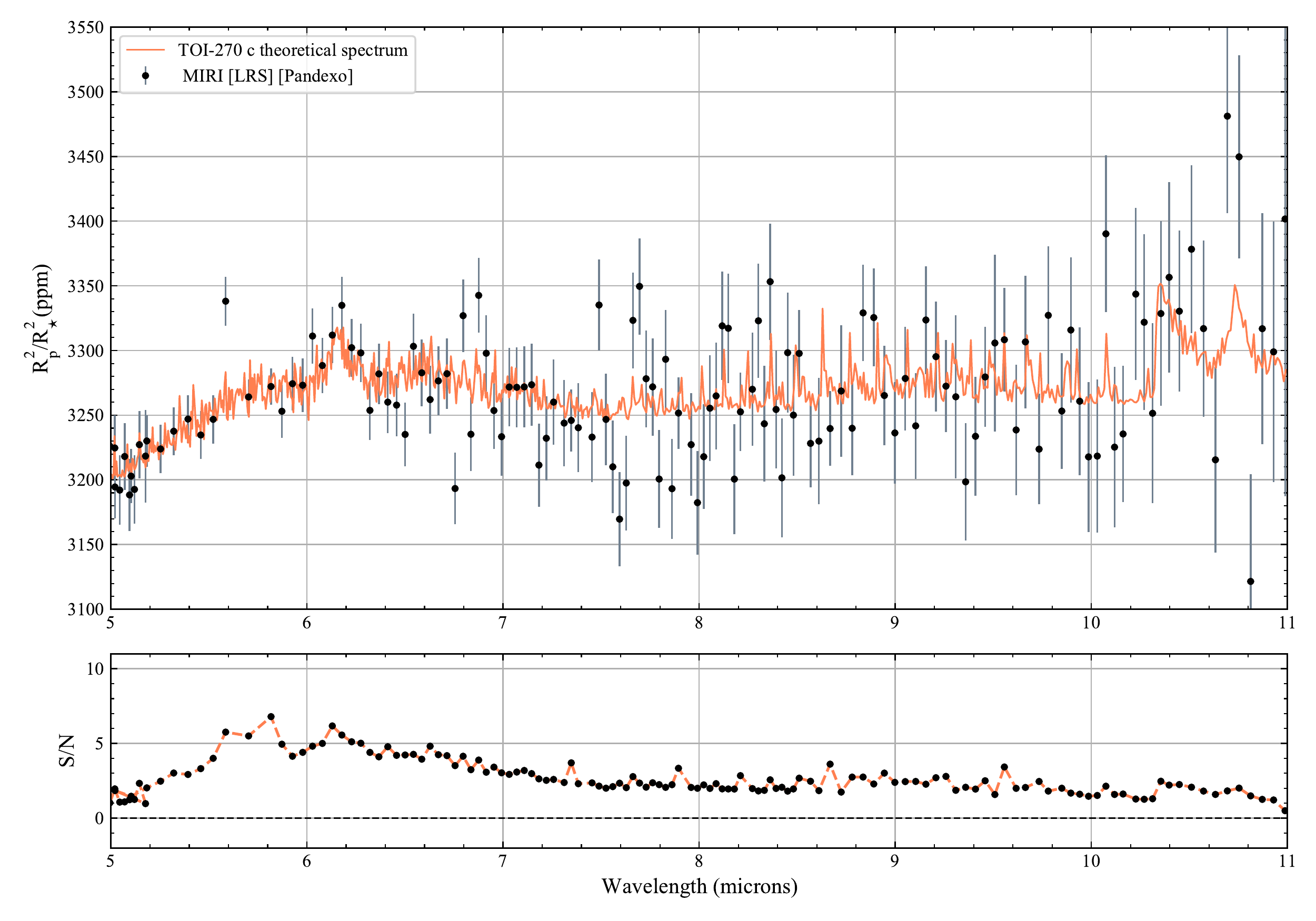}
    \caption{Simulated MIRI LRS transmission spectrum of TOI-270~c with 10 transits and corresponding S/N for the low MMW $\sim$ 4.6 case.}
    \label{fig:MIRI_transmission}
\end{figure}

 
We show a few examples of transmission spectra for TOI-270 c from Figure. \ref{fig:lowmmw_tranmission} to Figure. \ref{fig:MIRI_transmission}. As the S/N scales inversely with MMW,  we find that the ideal observing conditions are atmospheres with a low-MMW and clear atmosphere (Figure \ref{fig:lowmmw_tranmission}) as opposed to a high-MMW atmosphere that produces a weaker transmission signal and therefore lower S/N (Figure \ref{fig:highMMW}).

\par
For MIRI transmission spectroscopy, We find that the 6.1 micron NH$_{3}$ is more promising for detection with transmission spectroscopy with MIRI LRS than the 10.3--10.8 micron NH$_{3}$ feature (Figure \ref{fig:MIRI_transmission}) because of increasing noise towards longer wavelengths.

\begin{deluxetable}{ccccc}[htp!]
\tablenum{7}
\tablecaption{S/N of major near and mid-infrared NH$_{3}$ transmission features $\&$ ranking of targets \label{tab:ranking}}
\tablehead{
\colhead{Target} & \colhead{Ammonia Feature} & \colhead{S/N} & \colhead{$<${S/N}$>$} 
& \colhead{Ranking} \vspace{-2mm}\\ \colhead{}&\colhead{[$\mu$m]}& \colhead{[$\sigma$]}& \colhead{[$
\sigma$]}}
\startdata
\multirow{6}{*}{TOI-270~c}&1.5&5.0&\multirow{6}{*}{18.8}&\multirow{6}{*}{1}\\
&2.0&5.4\\
&2.3&5.1\\
&3.0&6.2\\
&6.1&5.5\\
&10.3--10.8&2.4\\
\hline
\multirow{6}{*}{LP 791-18~c}&1.5&5.4&\multirow{6}{*}{18.4}&\multirow{6}{*}{2}\\
&2.0&5.4\\
&2.3&5.7\\
&3.0&6.0\\
&6.1&5.3\\
&10.3--10.8&1.6\\
\hline
\multirow{6}{*}{TOI-270~d}&1.5&3.9&\multirow{6}{*}{12.0}&\multirow{6}{*}{3}\\
&2.0&4.2\\
&2.3&4.0\\
&3.0&4.3\\
&6.1&3.7\\
&10.3--10.8&1.7\\
\hline
\multirow{6}{*}{LHS 1140 b}&1.5&2.5&\multirow{6}{*}{6.6}&\multirow{6}{*}{4}\\
&2.0&2.6\\
&2.3&2.7\\
&3.0&2.6\\
&6.1&2.3\\
&10.3--10.8&1.0\\
\hline
\multirow{6}{*}{K2-3~c}&1.5&2.1&\multirow{6}{*}{5.7}&\multirow{6}{*}{5}\\
&2.0&2.2\\
&2.3&2.0\\
&3.0&2.4\\
&6.1&2.3\\
&10.3--10.8&0.9\\
\hline
\multirow{6}{*}{K2-18~b}&1.5&1.5&\multirow{6}{*}{3.7}&\multirow{6}{*}{6}\\
&2.0&1.6\\
&2.3&1.5\\
&3.0&1.7\\
&6.1&1.3\\
&10.3--10.8&0.5\\
\hline
\multirow{6}{*}{GJ 143 b \tablenotemark{a}}&1.5&--&\multirow{6}{*}{1.6}&\multirow{6}{*}{7}\\
&2.0&--\\
&2.3&--\\
&3.0&--\\
&6.1&1.5\\\
&10.3--10.8&0.7\\
\enddata
\tablenotetext{a}{GJ 143 b does not saturate MIRI LRS; however, there is saturation with the NIRSpec, and NIRISS instruments (Table \ref{tab:saturated_limit}), and therefore it is not a target for JWST based on its brightness and is ranked last even though it could be observed at 3 microns with the NIRCam grisms}
\end{deluxetable}
\par
 We also explore various conditions and atmospheric scenarios which could affect the detection level of ammonia in the atmosphere of our targets in the near-infrared: (1) varying concentration of ammonia in the atmosphere (\S \ref{sec:vary_ammonia}) (2) varying atmospheric composition (\S \ref{sec:vary_atmospheric});  and (3) varying cloud decks (\S \ref{sec:vary_clouds}). 
\subsection{Varying concentration of Ammonia}
\label{sec:vary_ammonia}
\begin{figure}[htp!]
    \centering
    \vspace{-0.20in}
    \includegraphics[width=1.1\columnwidth]{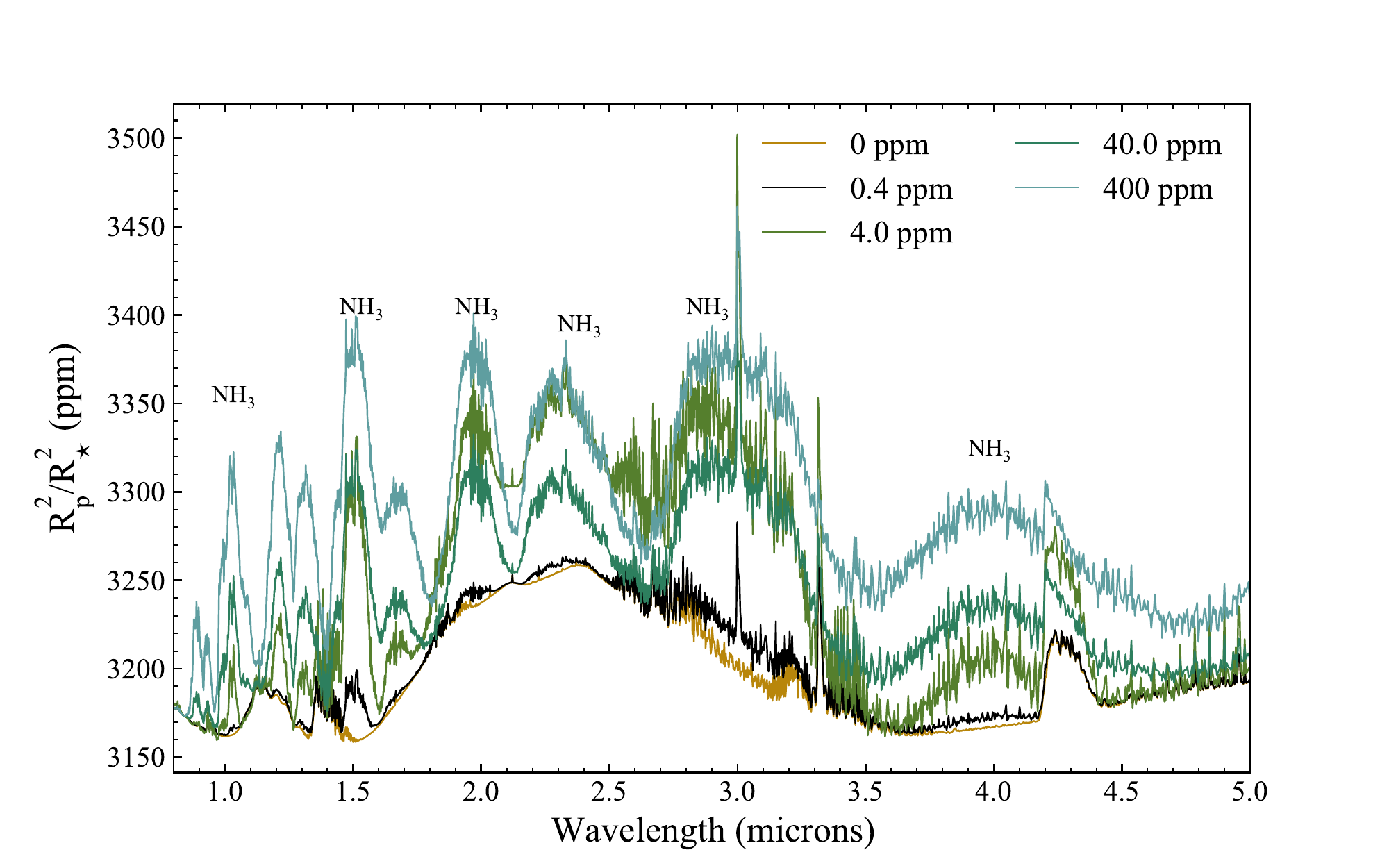}
    \caption{Theoretical transmission spectra of TOI-270~c with varying level of ammonia concentration: 0 ppm, 0.4 ppm, 4.0 ppm, 40 ppm, and 400 ppm.}
    \label{fig:ammonia_varying}
\end{figure}

Using TOI-270~c observations with NIRISS and NIRSpec as an example, we vary the amount of ammonia in the atmosphere (Figure \ref{fig:ammonia_varying}; Table \ref{tab:signaltonoise_varyingammonia}). Following \cite{seagarbiomass}, the concentration of ammonia in the atmosphere is proportional to the biomass on the surface.  A concentration of 11 ppm NH$_{3}$ in the atmosphere of a temperature planet \footnote{A cold-Haber world with 90$\%$ H$_{2}$:10$\%$ N$_{2}$ with T$_{eq}$ = 290 K and  1.75 R$_{\oplus}$} around a  weakly-active M-dwarf is produced by a biomass density of 1 gm$^{-2}$. In contrast, a quiet M-type star would need less biomass --- 1.4 $\times$ 10$^{-2}$ gm$^{-2}$, to produce the same 11 ppm concentration of NH$_{3}$ in the atmosphere (\citealt{Seager2013a}; \citealt{Seager2013b}).
 
\par
\begin{figure}
\vspace{-0.10in}
    \centering
    \includegraphics[width=1.1\columnwidth]{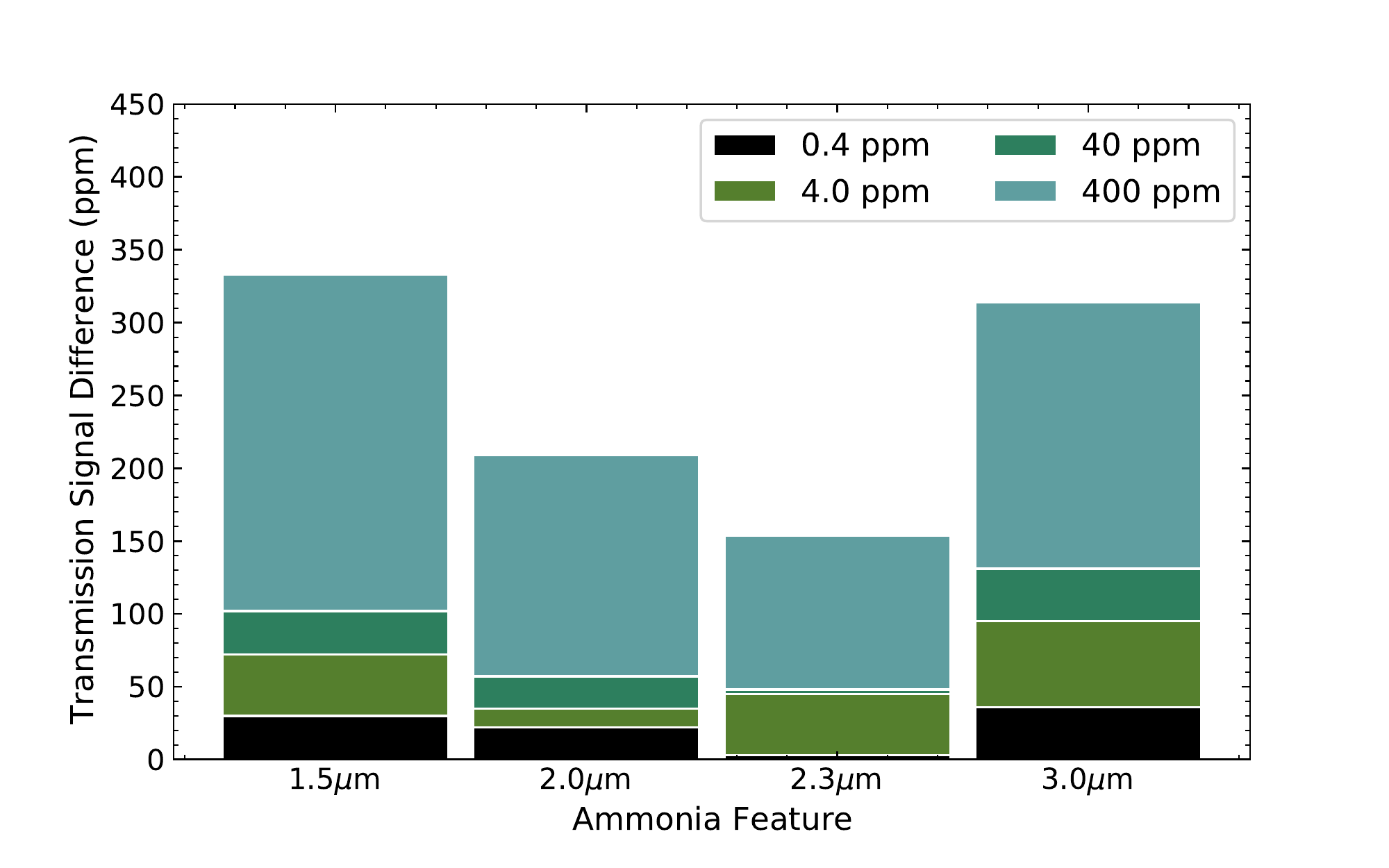}
    \caption{Transmission signal difference vs major transmission features for NH$_{3}$ (1.5, 2.0, 2.3, and 3.0 $\mu$m) for varying concentrations of ammonia in the atmosphere of TOI-270 c (0.4, 4.0, 40, and 400 ppm). A higher concentration of ammonia in the atmosphere (400 ppm) produces the highest change in the transmission signal strength as compared to 0 ppm of ammonia. }
    \label{fig:varying_bar_toi270c}
\end{figure}

We choose varying ammonia concentrations of 0.4, 4.0, 40, and 400 ppm. A range of 1--10 ppm concentration of NH$_{3}$ is reasonable for cold Haber Worlds. Our base simulations of detectability assume a 4.0 ppm concentration of ammonia, so we explore the effects of a varying ammonia concentration on the transmission signal on TOI-270~c. Figure \ref{fig:varying_bar_toi270c} shows that a higher concentration of ammonia produces higher signals for detection with the NIRSpec/G235M, NIRSpec/G395M, and NIRISS/SOSS modes (Table \ref{tab:signaltonoise_varyingammonia}). In the calculation of NH$_3$ signal as shown in Figure \ref{fig:varying_bar_toi270c}, we measure the transmission signal difference for the varying levels of ammonia in the atmosphere (0.4, 4.0, 40, and 400 ppm) for the 1.5, 2.0, 2.3, and 3.0 $\mu$m ammonia transmission features relative to the base transmission signal for a 0-ppm ammonia mixing ratio. 

\par

Figure. 3 in \cite{Seager2013b}, shows that vertical mixing can only increase NH$_{3}$ values higher up in the atmosphere, making it more detectable, so our simulation is a less optimistic case than the case that considers vertical mixing.

\begin{deluxetable*}{ccccc}[htp!]
\tablenum{8}
\tablecaption{Average S/N  and transmission signal difference of major NH$_{3}$ transmission features for TOI-270~c transmission spectroscopy for varying concentrations of ammonia }
\label{tab:signaltonoise_varyingammonia}
\tablehead{
\colhead{Concentration of NH$_{3}$} & \colhead{Ammonia Feature} & \colhead{S/N} & \colhead{$<$S/N$>$} &\colhead{Transmission Signal Difference}
\\ 
\colhead{}&\colhead{[$\mu$m]}& \colhead{[$\sigma$]} & \colhead{[$\sigma$]} &\colhead{[ppm]} }
\startdata
\multirow{4}{*}{0.4 ppm}&1.5&1.6&\multirow{4}{*}{8.5}&30\\
&2.0&5.1&&22\\
&2.3&5.2&&3\\
&3.0&4.1&&36\\
\hline
\multirow{4}{*}{4.0 ppm}&1.5&5.0&\multirow{4}{*}{10.9}&72\\
&2.0&5.4&&35\\
&2.3&5.1&&42\\
&3.0&6.2&&92\\
\hline
\multirow{4}{*}{40 ppm}&1.5&9.2&\multirow{4}{*}{16.0}&142\\
&2.0&7.2&&87\\
&2.3&6.0&&91\\
&3.0&9.0&&137\\
\hline
\multirow{4}{*}{400 ppm}&1.5&11.9&\multirow{4}{*}{19.5}&231\\
&2.0&8.8&&152\\
&2.3&6.7&&106\\
&3.0&10.5&&183\\
\enddata
\end{deluxetable*}
\vspace{-0.10in}
\subsection{Varying Atmospheric Composition}
\label{sec:vary_atmospheric}
Theoretical studies and observational works based on data from HST have explored the  atmospheric compositions for gas dwarf atmospheres (e.g. \citealt{elkins-seager2008}; \citealt{Miller_Ricci_2008}; \citealt{Seager_2010}; \citealt{Benneke_2012}).  Recently, HST observations showed that water vapor was present in the atmosphere which indicates a thick hydrogen-rich gas envelope for K2-18 b \citep{benneke2019, Madhusudhan2020}, and evidence of a thick atmosphere for LHS 1140  b~\citep{Edwards2021}. 
\par
\begin{figure}[htp!]
    \centering
    \vspace{-0.00in}
    \includegraphics[width=1.1\columnwidth]{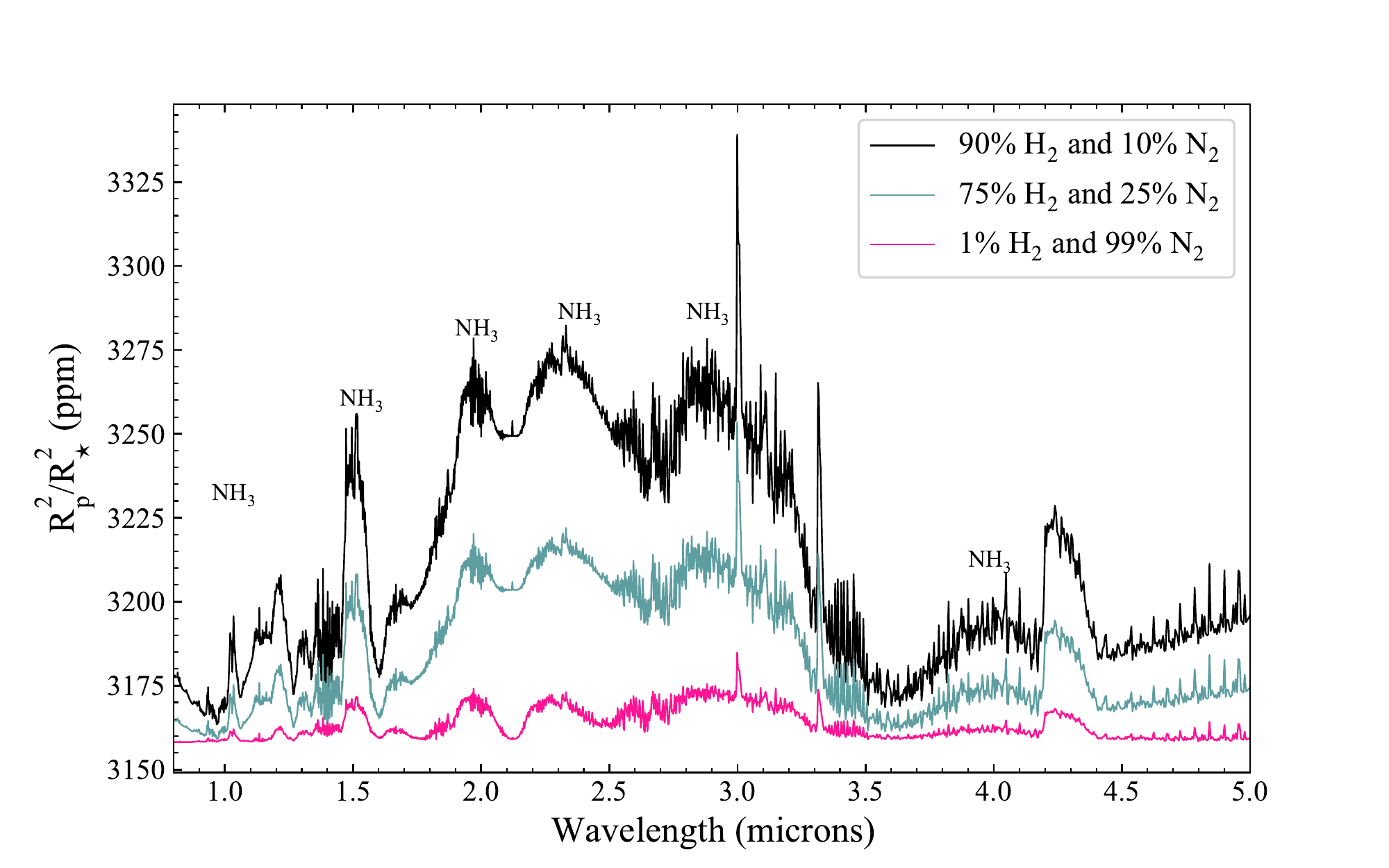}
    \caption{Modeled transmission spectra of TOI-270~c showing various atmospheric compositions compositions. The lines represent different hydrogen dominated scenarios: a H-rich atmosphere (90$\%$ H$_{2}$ and 10$\%$ $N_{2}$),
  a H-poor atmosphere (1$\%$ H$_{2}$ and 99$\%$ $N_{2}$),  and a H-intermediate atmosphere (75$\%$ H$_{2}$ and 25$\%$ $N_{2}$).}
    \label{fig:varying_atmospheric_comp}
\end{figure}

 
To explore the diversity of gas dwarf atmospheres, we follow \citealt{Chouqar2020} to consider the following cases: a hydrogen-rich atmosphere (90$\%$ H$_{2}$ and 10$\%$ $N_{2}$), a hydrogen-poor atmosphere (1$\%$ H$_{2}$ and 99$\%$ $N_{2}$) \footnote{The hydrogen-poor atmosphere is constructed following the same method as Section $\S$\ref{sec:atmospheric_composition} with  X$_\mathrm{{H}}$ = 0.01 $\&$ X$_\mathrm{{N}}$ = 0.99}, and a hydrogen-intermediate atmosphere (75$\%$ H$_{2}$ and 25$\%$ $N_{2}$), in order to determine the effects on the detection of ammonia. Figure \ref{fig:varying_atmospheric_comp} shows that the strength of ammonia features corresponds to the amount of hydrogen present in the atmosphere. Compared to a hydrogen-rich and hydrogen-intermediate atmosphere, a hydrogen-poor atmosphere has a weaker transmission signal and S/N - based on our  transmission detection metric. A more quantitative summary is given in Table \ref{tab:ranking_atm}.

\begin{deluxetable}{cccc}[htp!]
\tablenum{9}
\tablecaption{Average S/N of major NH$_{3}$ transmission features for TOI-270~c transmission spectroscopy for different atmospheric compositions }
\label{tab:ranking_atm}
\tablehead{
\colhead{TOI-270 c} & \colhead{Ammonia Feature} & \colhead{S/N} & \colhead{$<${S/N}$>$} 
\\ 
\colhead{}&\colhead{[$\mu$m]}& \colhead{[$\sigma$]}& \colhead{[$\sigma$]}
}
\startdata
\multirow{4}{*}{H-rich}&1.5&5.0&\multirow{4}{*}{10.9}\\
&2.0&5.4\\
&2.3&5.1\\
&3.0&6.2\\
\hline
\multirow{4}{*}{H-intermediate}&1.5&2.7&\multirow{4}{*}{5.9}\\
&2.0&2.9\\
&2.3&2.7\\
&3.0&3.4\\
\hline
\multirow{4}{*}{H-poor}&1.5&0.8&\multirow{4}{*}{1.6}\\
&2.0&0.7\\
&2.3&0.6\\
&3.0&1.0\\
\enddata
\end{deluxetable}
\vspace{-0.4in}
\subsection{Cloud Decks}
\label{sec:vary_clouds}
One of the major challenges for the search of NH$_{3}$ in gas dwarfs is the presence of clouds; which have been shown to mask transmission features (\citealt{Helling2019}; \citealt{Barstow2021}). Clouds have been shown to be present in the atmospheres of gas dwarfs (\citealt{Kreidberg2014}; \citealt{benneke2019}). For example, GJ 1214 b,  has shown a flat transmission spectrum due to the effects of high-altitude clouds (\citealt{Berta2012}; \citealt{Kreidberg2014}). 
\par
\begin{figure}[htp!]
    \centering 
    \includegraphics[width=1.1\columnwidth]{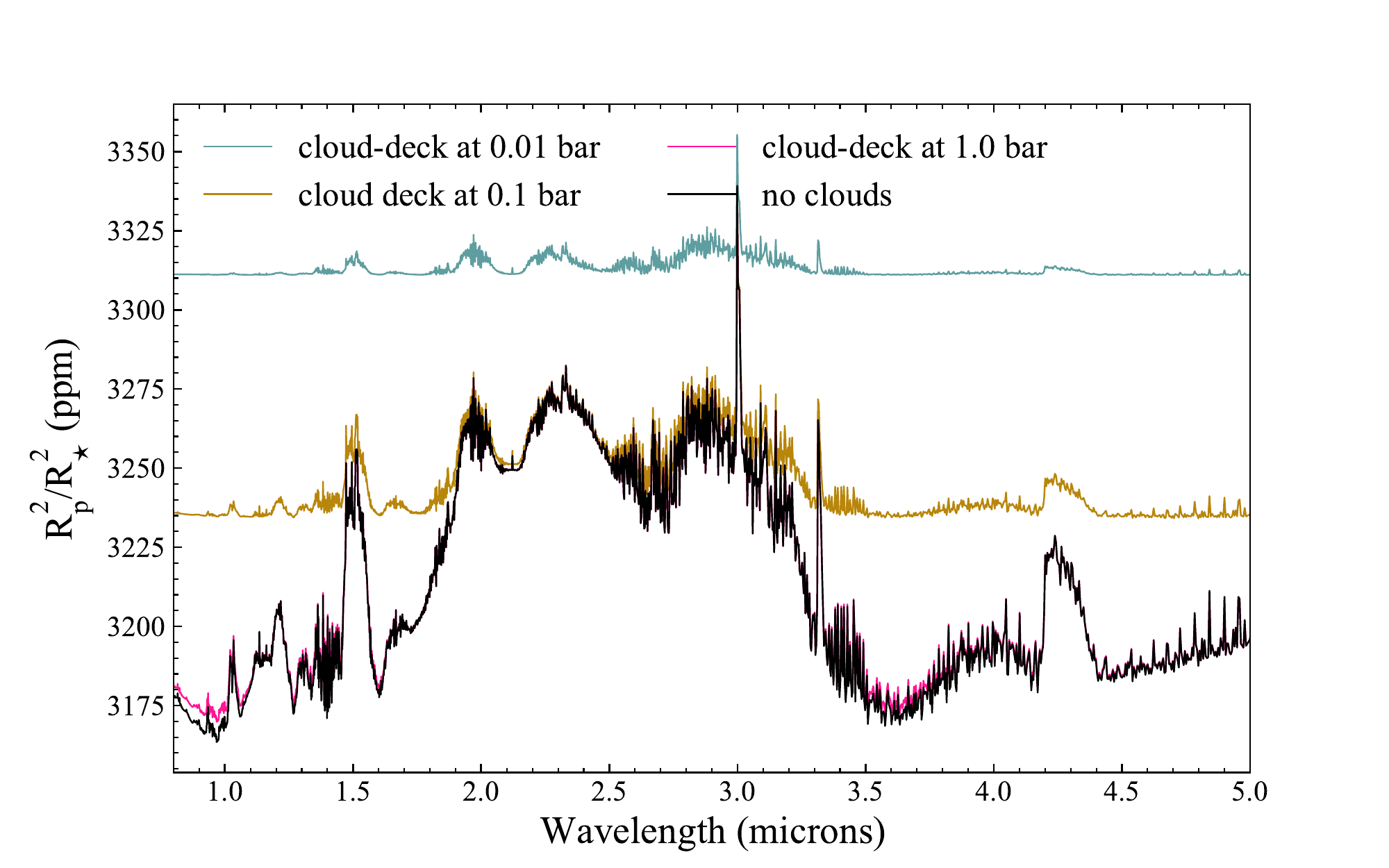}
    \caption{Transmission spectra of TOI-270 c shown with varying cloud deck structures at 0.01 bar (blue), 0.1 bar (gold) and 1 bar (pink). The varying ammonia feature are muted with a decreasing cloud deck pressure and are almost a flat line with a 0.01-bar cloud deck.}
    \label{fig:clouds_decks}
\end{figure}

We use \texttt{petitRADTRANS} to model the effect of varying levels of cloud deck structures in the atmospheres (1 bar, 0.1 bar,  and 0.01 bar). We find that a decreasing cloud deck pressure (i.e., increasing height of clouds) masks the NH$_{3}$ atmospheric features to a near flat continuum and affects the detectability of major NH$_{3}$ transmission feature (Figure \ref{fig:clouds_decks}). A more quantitative summary is given in Table \ref{tab:ranking_clouds}. \footnote{The S/N [$\sigma$] values are calculated using our transmission detection metric (Equation \ref{eq:S/N})} 

\par
We employ the use of a contribution function  which indicates locations of a  cloud deck layer in the atmosphere that transmission features are produced from and where spectral features begin to become muted. 

\begin{figure}
    \centering
    \includegraphics[width=1.1\columnwidth]{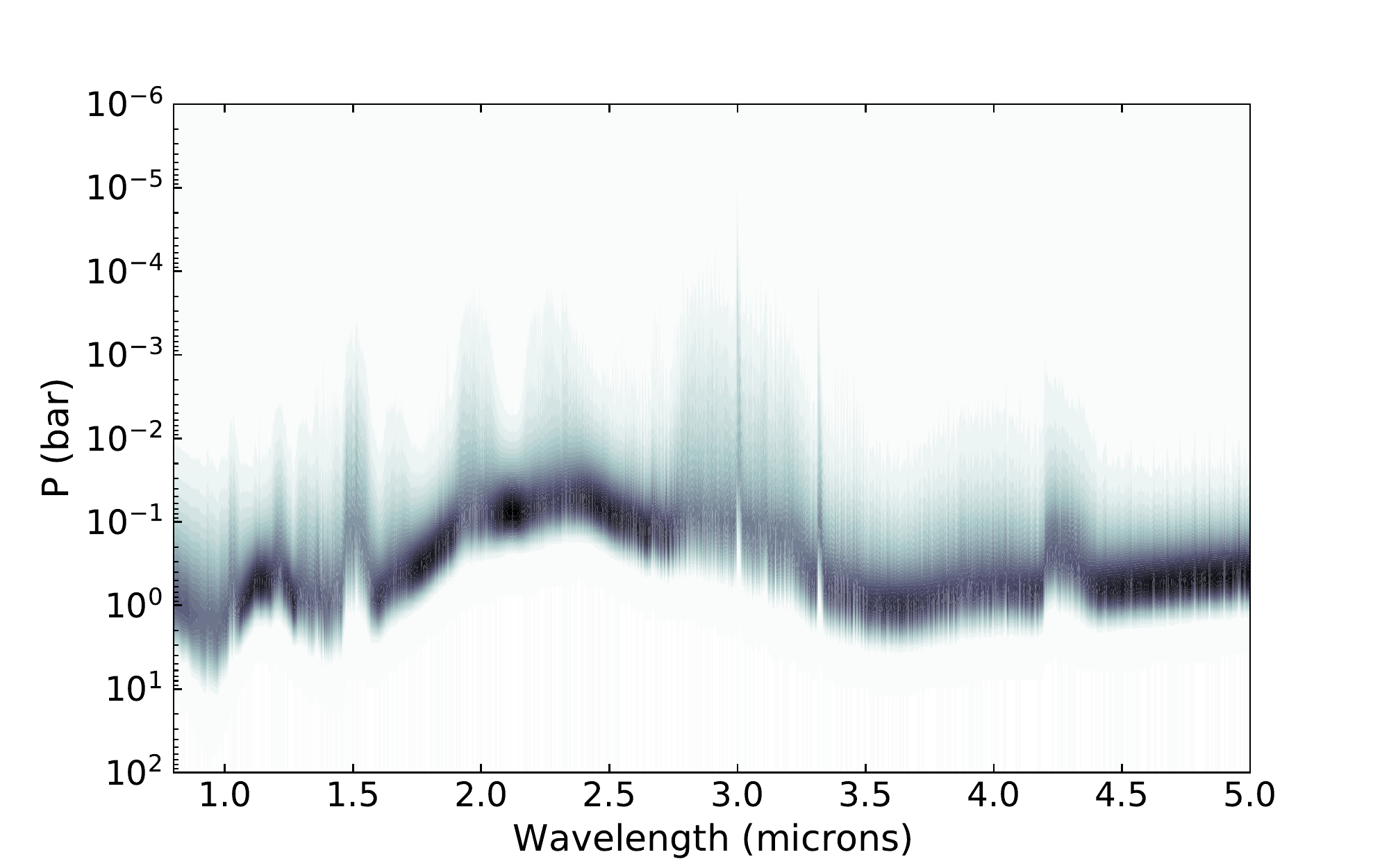}
    \caption{Contribution function vs wavelength for TOI-270 c. Maximum contributions are shown in black. Primarily the transmission features are produced at pressures above 10$^{-2}$ bar.}
    \label{fig:contribution_TOI-270c}
\end{figure}

We find that primarily the transmission features are produced at pressures higher than 10$^{-2}$ bar in the atmosphere. Therefore, a cloud deck at the pressure level (and below) starts to mute the spectral features (Figure \ref{fig:contribution_TOI-270c}). 

\begin{deluxetable}{cccc}[htp!]
\tablenum{10}
\tablecaption{Average S/N of major NH$_{3}$ transmission features for TOI-270~c transmission spectroscopy for varying cloud decks with a H-rich atmosphere.  \label{tab:ranking_clouds}}
\tablehead{
\colhead{TOI-270 c} & \colhead{Ammonia Feature} & \colhead{S/N} & \colhead{$<${S/N}$>$} 
\\ 
\colhead{}&\colhead{[$\mu$m]}& \colhead{[$\sigma$]}& \colhead{[$\sigma$]}
}
\startdata
\multirow{4}{*}{Cloud deck at 0.01 bar}&1.5&0.4&\multirow{4}{*}{1.2}\\
&2.0&0.3&\\
&2.3&0.3\\
&3.0&1.0&\\
\hline
\multirow{4}{*}{Cloud deck at 0.1 bar}&1.5&2.0&\multirow{4}{*}{4.8}\\
&2.0&2.0&\\
&2.3&2.2\\
&3.0&3.2&\\
\hline
\multirow{4}{*}{Cloud deck at 1.0 bar}&1.5&4.9&\multirow{4}{*}{10.8}\\
&2.0&5.3&\\
&2.3&5.0\\
&3.0&6.1&\\
\enddata
\end{deluxetable}

\par
Hazes can impact spectra and produce flatter transmission spectra similar to clouds (\citealt{Marley2013}; \citealt{wunderlich2020}). Our grey cloud treatment results in the same behavior as hazes in the near infrared. Super Rayleigh slope due to hazes~\citep{Ohno2020} primarily impact the bluer optical wavelengths - so we do not explore the effects in this study.

\section{Atmospheric Retrieval Examples}
\label{sec:atmospheric_retrievals}

In previous sections, we provide an SNR-based metric to quantify the detectability of NH$_3$. The metric guides us in prioritizing targets and determining conditions under which the detection is plausible. In this section, we provide a few examples on how NH$_3$ can be detected and the abundance constrained under optimal conditions, e.g., a cloud-free atmosphere with low MMW for TOI-270 c, one of the most promising targets in our sample based on the SNR metric. 

A full exploration of parameter space will be conducted in a future paper to determine the threshold for detecting and constraining NH$_3$ abundance. The purpose here is to (1) demonstrate that NH$_3$ and H$_2$O abundance can be retrieved independently despite their overlapping wavelengths in absorption; and (2) understand the impact of cloud on the retrieval precision. 

\subsection{Setting Up the Retrieval}
We use the simulated JWST data for TOI-270 c as the input. To model the simulated data, we use \textsc{petitRadTRANS}~\citep{molliere2019} with the following free parameters: surface gravity, planet radius, temperature for the isothermal atmoshpere, cloud deck pressure, and mass mixing ratios for different species that are being considered. 
In a Bayesian framework, we use \textsc{PyMultiNest}~\citep{Buchner2014} to sample the posteriors. The priors are listed in Table \ref{tab:prior}. The likelihood function is $\sum\exp[{-(\mathcal{D}-\mathcal{M})^2/\mathcal{E}^2}]$/2, where $\mathcal{D}$ is data, $\mathcal{M}$ is model, and $\mathcal{E}$ is the error term. The input values of the retrieval tests can be found in Table \ref{tab:prior}. For \texttt{PyMultiNest}, we use 2000 live points.

\begin{deluxetable*}{lccccccc}
\tablenum{11}
\label{tab:prior}
\tablecaption{Parameters used in retrieval, their priors, input and retrieved values}
\tablehead{
\colhead{\textbf{Parameter}} &
\colhead{\textbf{Unit}} &
\colhead{\textbf{Type}} &
\colhead{\textbf{Lower}} &
\colhead{\textbf{Upper}} &
\colhead{\textbf{Input}} & 
\multicolumn{2}{c}{\textbf{Retrieved}} \\
\colhead{\textbf{}} &
\colhead{\textbf{}} &
\colhead{\textbf{}} &
\colhead{\textbf{}} &
\colhead{\textbf{}} &
\colhead{\textbf{}} & 
\colhead{\textbf{Fixed}} & 
\colhead{\textbf{Free}} \\
}




\startdata
Surface gravity ($\log$g)                   &  cgs                 &   Uniform        &  2.0        &   5.0   & 3.0395  & $      3.20^{     +0.19}_{     -0.34}$ & $      3.12^{     +0.24}_{     -0.35}$ \\
Planet radius (R$_P$)                       &  R$_{\rm{Jupiter}}$  &   Uniform        &   0.2      &   0.5    & 0.208 & $   0.20802^{  +0.00035}_{  -0.00013}$ & $   0.20805^{  +0.00035}_{  -0.00017}$\\
Temperature (T$_{\rm{iso}}$))           &  K                 &   Log-uniform       &  10        &   3300     & 400 & $436^{     +53}_{     -47}$ & $436^{     +53}_{     -38}$\\
Cloud pressure ($\log$(P$_{\rm{cloud}}$))           &  bar                 &   Log-uniform       &  -6        &   6    & 5.05 & fixed & $      2.46^{     +2.38}_{     -2.18}$ \\
H$_2$O Mixing Ratio ($\log$(mr$_{\rm{H}_2\rm{O}}$)) &  \nodata             &   Log-uniform      &  -10       &   0   & -5.44  & $     -5.39^{     +0.31}_{     -0.94}$ & $     -5.67^{     +0.44}_{     -0.99}$\\
CO Mixing Ratio ($\log$(mr$_{\rm{C}\rm{O}}$))       &  \nodata             &   Log-uniform       &   -10     &   0   & -8.25 & fixed & $     -8.35^{     +1.33}_{     -1.09}$ \\
CO$_2$ Mixing Ratio ($\log$(mr$_{\rm{C}\rm{O}_2}$)) &  \nodata             &   Log-uniform       &  -10       &   0   & -7.55 & fixed & $     -8.14^{     +0.66}_{     -0.99}$ \\
CH$_4$ Mixing Ratio ($\log$(mr$_{\rm{C}\rm{H}_4}$)) &  \nodata             &   Log-uniform       &  -10       &   0   & -6.99 & $     -7.20^{     +0.56}_{     -1.54}$ & $     -7.34^{     +0.62}_{     -1.31}$ \\
OH Mixing Ratio ($\log$(mr$_{\rm{O}\rm{H}}$)) &  \nodata             &   Log-uniform       &  -10       &   0   & -14.47 & fixed & $     -6.11^{     +1.29}_{     -2.38}$ \\
NH$_3$ Mixing Ratio ($\log$(mr$_{\rm{N}\rm{H}_3}$)) &  \nodata             &   Log-uniform       &  -10       &   0   & -4.86 & $     -4.76^{     +0.23}_{     -0.46}$ & $     -4.89^{     +0.31}_{     -0.51}$ \\
H$_2$ Mixing Ratio ($\log$(mr$_{\rm{H}_2}$)) &  \nodata             &   Log-uniform       &  -10       &   0   & -0.44 & $     -0.29^{     +0.20}_{     -0.42}$ & $     -0.37^{     +0.26}_{     -0.44}$ \\
HCN Mixing Ratio ($\log$(mr$_{\rm{H}\rm{C}\rm{N}}$)) &  \nodata             &   Log-uniform       &  -10       &   0   & -8.27 & fixed & $     -8.05^{     +1.38}_{     -1.28}$ \\
Wavelength shift ($\Delta_\lambda$)         &  $\mu$m              &          Uniform                &   -0.1     &   0.1  & 0.0 & fixed  & $   0.00066^{  +0.00198}_{  -0.00281}$  \\
\enddata

\end{deluxetable*}



\subsection{Fixing Cloud Deck and Other Minor Species}
In this case, we use the simulated data for the cloud-free low-MMW case for TOI-270 c (Figure. \ref{fig:lowmmw_tranmission}). In the retrieval, we assume the cloud deck at $10^{5}$ bar, which is well below the pressure range that contributes to the absorption (i.e., $10^{-3}-10$ bar). This is therefore an a priori cloudy-free case for the retrieval. Furthermore, we fix the abundance for species whose mass mixing ratios are below $5\times10^{-8}$ for the following reasons. First, these species have low abundances and may not be practically detected given the JWST data quality. This will be shown in the next example. Second, we want to focus on NH$_3$ and H$_2$O in this example and check if the two species can be reasonably measured despite overlapping absorption wavelengths regions.

\begin{figure*}[htp!]
    \centering
    \includegraphics[width=1.9\columnwidth]{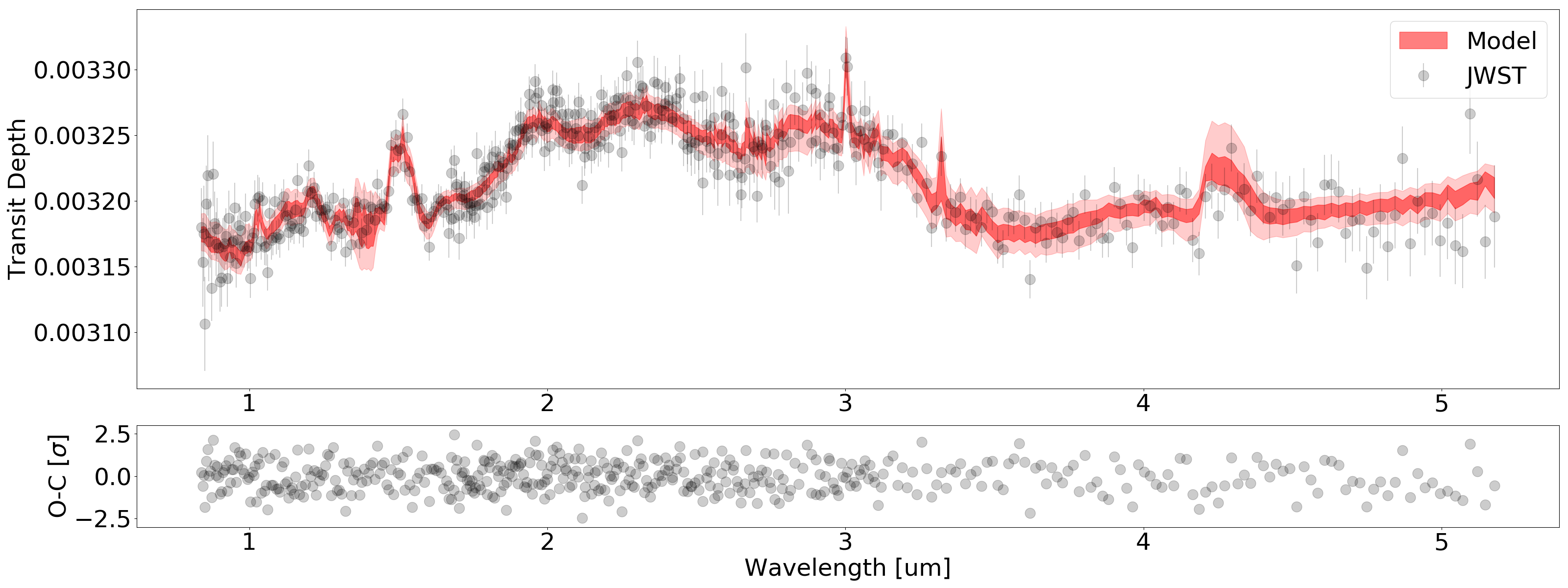}\\
    \includegraphics[width=1.9\columnwidth]{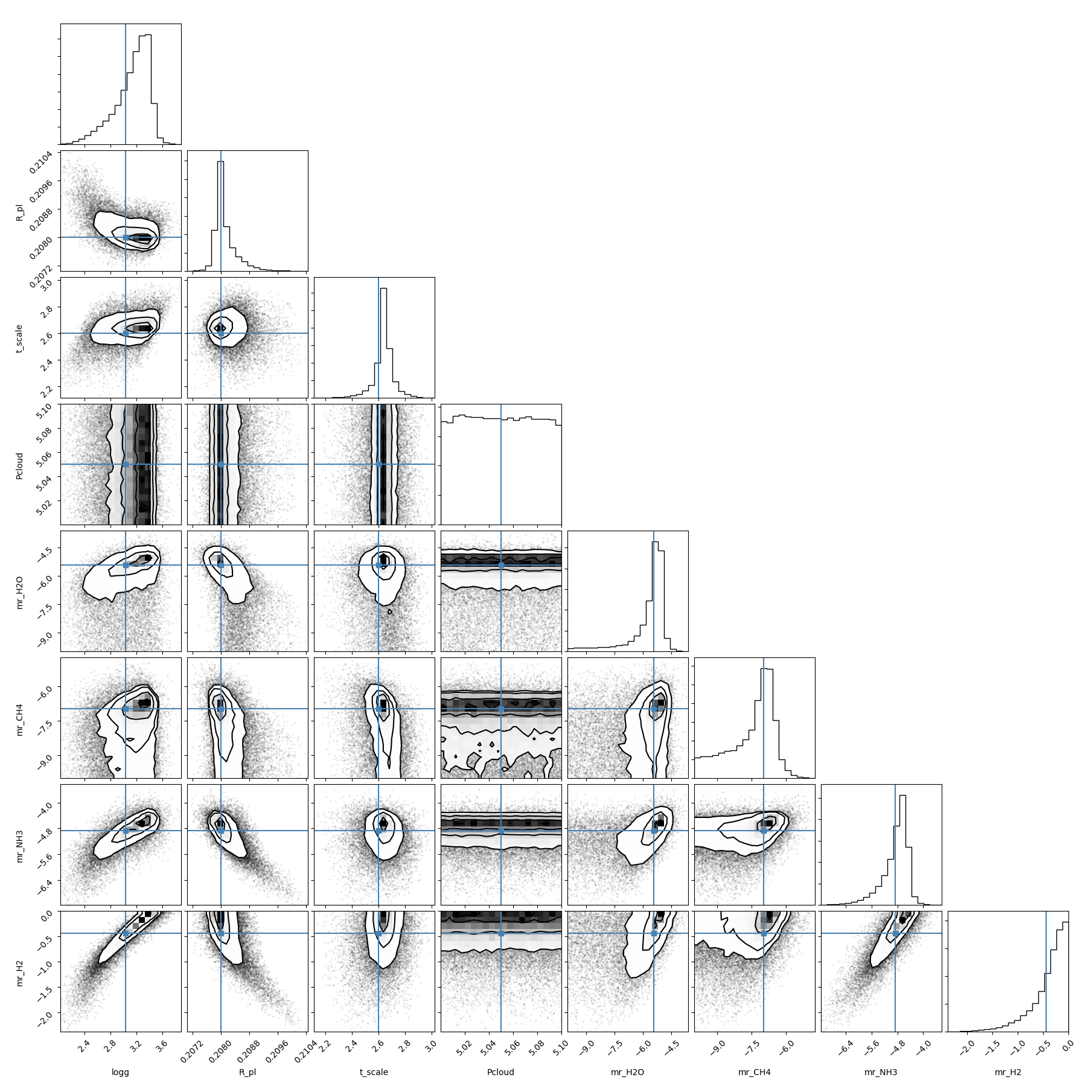}\\
    \caption{Top: Simulated JWST data vs. the retrieved model, and the O-C plot that shows the residual. Bottom: Corner plot for the parameters that are used in the retrieval along with true values that are used in generating the JWST data. Contours are shown at 0.5, 1, 1.5, and 2-$\sigma$.}
    \label{fig:retrieval_fixed_cloud}
\end{figure*}

As shown in Fig. \ref{fig:retrieval_fixed_cloud}, NH$_3$ and H$_2$O can be detected in our retrieval, and their abundances are within $1-\sigma$ from the input values. The comparison between simulated data points and retrieved spectra shows a good agreement, although the $2-\sigma$ region for the modeled spectra does not cover the majority of the data points. This is particularly the case for the 2.0 and 2.3 NH$_3$ features. We attribute this issue to the limitation of our modeling software in generating arbitrary shapes of spectra, but point out that the limitation of modeling spectra can be properly accounted for by adding a Gaussian Process component in the retrieval process~\citep[e.g.,][]{Wang2020b}.

We have two ways of quantifying the detection significance. First, given the $\sim$50,000 posterior samples and that no point falls into the [10$^{-10}$-10$^{-9}$] mixing ratio bin, we have a lower limit of $4.2-\sigma$ detection for NH$_3$ and H$_2$O. Second, using the retrieved NH$_3$ abundance of -4.76 dex and a $1-\sigma$ error bar of 0.46 dex (see Table \ref{tab:prior}), the distance to the lower edge of the prior -10 dex is $11.4-\sigma$. This is consistent with values reported in Table \ref{tab:ranking}: the quadrature summation of S/N for the 1.5, 2.0, 2.3, and 3.0 NH$_3$ features is $10.9-\sigma$.

\subsection{Cloud Deck as a Free Parameter}

We then run a retrieval analysis on the full parameter set that includes (1) the cloud deck pressure; and (2) all minor species with mass mixing ratio lower than $5\times10^{-8}$. This is to demonstrate that the retrieval works with the full parameter set and returns a reasonable result when compared to the input parameters. 

We have an upper limit for the cloud deck pressure at $\sim$3 bar. {{This is consistent with the contribution function (Figure. \ref{fig:contribution_TOI-270c}) which shows that the majority of planet transmission signal comes from atmospheric layers with pressures lower than 3 bar. Note that the prior for the cloud deck pressure covers a range from $10^{-6}$ to $10^{6}$ bar.{The comparison between the data and the retrieved modeled spectra is similar to Figure. \ref{fig:retrieval_fixed_cloud} and we therefore do not include a comparison figure for this case.}}}

\begin{figure*}[htp!]
    \centering
    \includegraphics[width=2.0\columnwidth]{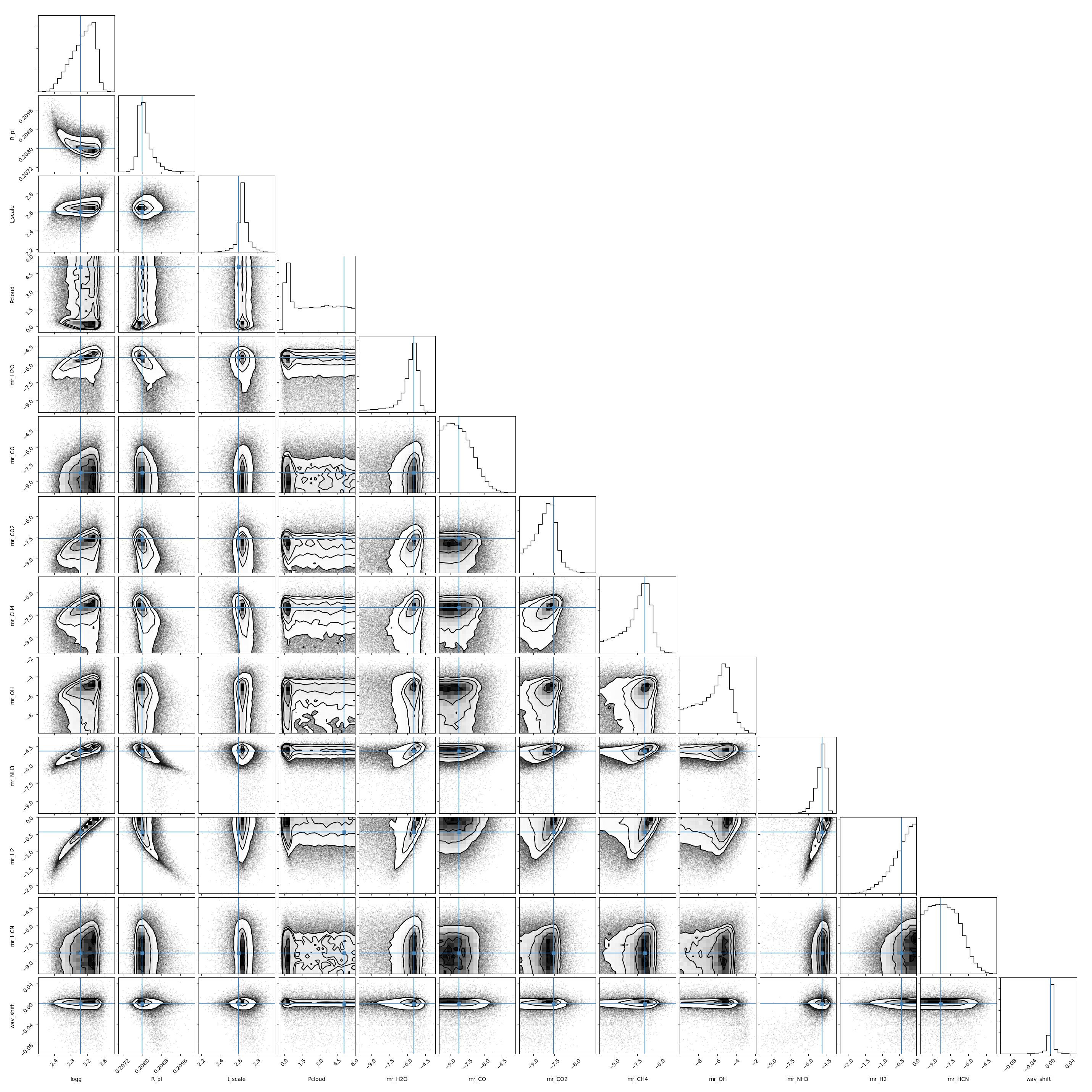}
    \caption{Corner plot for the full parameter set that is used in the retrieval along with true values that are used in generating the JWST data.}
    \label{fig:retrieval_vary_cloud}
\end{figure*}

\section{Discussion}
\label{sec:discussions}
\subsection{Comparing to 
\citealt{Chouqar2020}}
\cite{Chouqar2020} performed a comprehensive study on the properties of the TOI-270 system including the detectability of an atmosphere and individual molecules using the NIRISS (SOSS), and NIRSpec/G395M modes for transmission spectroscopy in the near-infrared.  

\par
To calculate the total expected signal-to-noise and number of transits to detect spectral features so that 
a $<$S/N$>$ $\geq$ 5, the approach presented in \cite{Yaeger2019} is utilized.  Similar to \cite{Yaeger2019}, a S/N scaling relation is developed by running the \texttt{PandExo} JWST noise model across a grid with transits ranging from 1 - 100. A S/N is determined based on the difference between the model spectrum and featureless fiducial spectrum. This is the basis of the approach used, but a more detailed description of the approach is available in \cite{Yaeger2019} and \cite{Chouqar2020}.
\par

We find that for both instrument modes there are discrepancies between our results  for TOI-270~c, also seen for TOI-270~d. Similarly to our work, \cite{Chouqar2020} utilize \texttt{petitRADTRANS} to generate the spectra, for a clear hydrogen-rich atmosphere; however, they used a MMW = 2.39 (X$_{H}$= 0.9), as opposed to our hydrogen-rich atmosphere  with a MMW = 4.55. As a result, their simulated 2.0 $\mu$m NH$_{3}$ feature for TOI-270~c has a $\sim$ 200 ppm signal from the baseline and our spectrum of TOI-270~c has a $\sim$ 100 ppm signal. 
\par
 For TOI-270~c with a  H-rich atmosphere, they find that the NIRISS (SOSS) instrument requires only one transit to  detect ammonia\footnote{\citet{Chouqar2020} consider the 2.0 $\mu$m ammonia feature} with a SNR = 18. We attempt to replicate their results using their MMW value, R$\sim$10, and the number of transits set to 1. For the same 2.0 $\mu$m feature using NIRISS (SOSS) we find a S/N= 12.5.

\par
They also find that with NIRSpec/G395M for 1 transit, the ammonia feature (we assume they are referring to the 3.0 $\mu$m feature, based on their Figure 5) has a S/N of 5.0. For NIRSpec/G395M we find a S/N = 4.0 for the 3.0$\mu$m feature. Therefore, our NIRSpec/G395M simulation is roughly consistent with the result in \cite{Chouqar2020}.

\subsection{Comparing to \citealt{wunderlich2020}}
 \cite{wunderlich2020} performed an assessment of the detectability of biosignatures for LHS 1140 b with NIRSpec/PRISM. For the purpose of comparison, we look specifically at the  3.0 $\mu$m NH$_{3}$ feature they considered for their study.

\par

In their study, a S/N = 5 is utilized to determine whether or not a spectral feature is detectable. The method employed involves the subtraction of the full transmission spectra  with all included absorption species from the spectrum excluding contribution from individual species. The S/N ratio determination is based on \citealt{Wunderlich2019}. 
\par
Differing from \texttt{petitRADTRANS}, which is based on a radiative transfer model, the atmosphere of LHS 1140 b from \cite{wunderlich2020} is built using the radiative-convective  photochemistry-climate coupled model, \texttt{1D - TERRA}.  The \texttt{1D - TERRA} model has inputs of the following: P-T profiles, initial compositions, stellar spectrum, and ion-pair production rates.  Detailed schematics of the full model description is shown in Figure 1 of \cite{Scheucher2020}.

Compared to our study, \cite{wunderlich2020} additionally varies the concentration of CH$_{4}$ with the \enquote{low CH$_{4}$} scenario assuming a VMR of $\sim$ 1$\times$10$^{-6}$, which is  about two orders of magnitude higher than our VMR for CH$_{4}$ of $\sim$ 2.9$\times$10$^{-8}$  for our low MMW 90$\%$ H$_{2}$ model atmosphere.

They find that with NIRSpec/PRISM - to detect the 3.0 $\mu$m NH$_{3}$, the required time for a H$_{2}$-dominated atmosphere with a \enquote{low CH$_{4}$} scenario,  the minimum number of transits would be $\sim$ 30 transits, to achieve a S/N of 5. Overall, \cite{wunderlich2020} find that to detect NH$_{3}$ the required time would be between 10--50 transits ($\sim$ 40--200) hours of observing time assuming non-cloudy conditions. 

\par
 to \cite{wunderlich2020} we do not consider the NIRSpec/PRISM as this mode saturates the detector for LHS 1140 b (Table \ref{tab:saturated_limit}; \cite{wunderlich2020}); instead we employed the use of NIRSpec/G235M, NIRSpec/G395M, and NIRISS (SOSS) modes. For LHS 1140~b based on our defined detection metric, we find that using NIRISS (SOSS) for the 3.0 $\mu$m NH$_{3}$, with 10 transits we would find a 3.1$\sigma$ detection given clear atmosphere conditions in a H-rich atmosphere, which corresponds to $\sim$ 60 hours of observing time. For 30 transits  ($\sim$ 180) hours of observing time) we find a S/N $\sim$ 4.4$\sigma$. This is qualitatively consistent with their conclusion that NH$_{3}$  can be detected in 10--50 transits, and the detection of NH$_{3}$ can be made in the presence of higher concentration of CH$_{4}$.

\subsection{False Positives for Biotic NH$_{3}$}
 Ammonia has been proposed as a biosignature in hydrogen-dominated atmospheres, however, it is not immune to false positives~\citep{Seager2013b,Catling2018}. \citet{Seager2013b} defined three major factors that can cause ammonia to be produced abiotically in these atmospheres: (1) a rocky world with a surface temperature of $\sim$ 820 K \footnote{At high surface temperatures, ammonia can be produced by the traditional Haber process from an iron surface}, (2) the natural production of NH$_{3}$ in the atmospheres of mini-Neptunes, (3) planets that have outgassed NH$_{3}$ during evolution. \citet{Seager2013b} notes that targets have to be evaluated on a case-by-case basis. Additionally,  according to the thesis work by Evan Sneed\footnote{\url{https://zenodo.org/record/4015708\#.YJXiGy1h124}}, another cause of false positives for NH$_{3}$ include comet collisions that contain inorganic ammonia ice.  
 
\subsection{Other Factors in Prioritizing Targets}

Other factors beyond the S/N of NH$_{3}$ for our targets should be considered when prioritizing targets for JWST time. These factors include precise mass and radius measurements. 

\par
For example, only K2-18 b~\citep{benneke2019} and LHS 1140 b~\citep{Dittmann2017, lillobox2020} have better than $\sim$15\% and $\sim$3\% precision in mass and radius measurements. This allows for a proper modeling of the planet interior~\citep{lillobox2020} and atmosphere composition~\citep{Madhusudhan2020}, which is essential to interpret the results of detection and non-detection.

\section{Conclusion}
\label{sec:conclusions}

 We modeled seven promising  gas dwarfs for the detection of the potential biosignature ammonia using the MIRI, NIRSpec, and NIRISS instruments on the upcoming JWST mission: GJ 143 b, TOI-270~c, TOI-270~d, K2-18 b, K2-3~c, LHS 1140~b, and LP 791-18~c. 
\par

MIRI LRS has an systemic noise limit of $\sim$ 12.6 ppm for 10 eclipses, where the 10.3--10.8 $\mu$m NH$_{3}$  feature for the majority of our targets is not detectable due to the  limitation by this systematic noise limit.

The most promising targets for emission spectroscopy with MIRI LRS is LP 791-18~c in terms of expected emission spectroscopy signal (Figure \ref{fig:fpfstar}). However, in practice, even with 10 transits (i.e $\sim$ 40 hours of observing time) we cannot realistically detect the signal (Figure \ref{fig:emission_spectra_lp}) due to large photon noise.

\par
We defined a detection metric for transmission spectroscopy and utilize the following modes to perform JWST simulations using a baseline of 10 transits for non-cloudy and low MMW atmospheres: NIRSpec/G395M, NIRSpec/G235M, NIRISS (SOSS), and MIRI LRS. We compile a ranking list for observing targets with JWST based on the S/N detection metric of  six major NH$_{3}$ features (1.5, 2.0, 2.3, 3.0, 6.1, and 10-3--10.8 $\mu$m). The rank list follows as such: TOI-270~c, LP 791-18~c TOI-270~d, LHS 1140~b, K2-3~c, K2-18~b, and GJ 143~b. TOI-270~c, is ranked first as it has the highest average S/N and GJ 143~b is ranked last as the host star saturates the majority of the chosen observing modes.
\par
We also test the capabilities of  transmission spectroscopy with MIRI LRS to detect the 6.1 and 10.3--10.8 $\mu$m NH$_{3}$ features and find that TOI-270~c has the highest $<$S/N$>$ and is best suitable to detect the 6.1 micron feature, and overall we find that the 6.1 micron feature is more suitable for detection than the 10.3--10.8 micron NH$_{3}$ feature for transmission spectroscopy with MIRI LRS.

Using TOI-270~c as an example, we test a variety of scenarios to determine the effect of detectability of ammonia: varying concentration of ammonia (\S \ref{sec:vary_ammonia}), varying atmospheric composition  (\S \ref{sec:vary_atmospheric}), and including effects of cloud decks (\S \ref{sec:vary_clouds}).

For a baseline of 10 transits, we find that a higher concentration of ammonia (400 ppm) in the atmospheres produces a higher transmission signal difference. Similarly, we model the effects of a varying hydrogen composition for TOI-270~c, and find that a H-rich (90$\%$ hydrogen based atmosphere) produces a higher averaged S/N detection, about an factor of 10, compared to a  H-poor (1$\%$ hydrogen based atmosphere). Lastly,  in the presence of cloud decks, the average S/N detection of ammonia  decreases to 1.2$\sigma$ and 4.8$\sigma$ from 10.9$\sigma$ from the cloud deck of 0.01 bar and 0.1 bar respectively.

We provide examples of atmospheric retrieval (\S \ref{sec:atmospheric_retrievals}) and show that NH$_3$ and H$_2$O can be detected in amenable conditions, i.e., a cloud-free atmosphere with a low MMW. For example, based on the posterior distribution of NH$_3$ in Figure. \ref{fig:retrieval_fixed_cloud}, the detection significance is $\sim$11-$\sigma$, which is roughly consistent with the S/N value that is reported in Table \ref{tab:ranking} for the four NH$_3$ feature from 1.5 to 3.0 $\mu$m.  The comparison provides corroborative evidence for the validity of our method of calculating S/N and the retrieval code. The retrieval can also constrain the pressure level of a cloud deck. 

This work demonstrates that JWST will provide unprecedented wavelength coverage and light collecting area for atmospheric studies of gas dwarfs and their potential biosignatures. 

\facilities{Exoplanet Archive}
\software{\texttt{Pandexo} \citep{batalha2017}, \texttt{petitRADTRANS} \citep{molliere2019}} 

\noindent
{\bf{Acknowledgments}} We thank the anonymous referee for their time providing helpful comments which improved the quality of this paper.
This research has made use of the NASA Exoplanet Archive, which is operated by the California Institute of Technology, under contract with the National Aeronautics and Space Administration under the Exoplanet Exploration Program.
\par
NASA's Astrophysics Data System Bibliographic Services together with the VizieR catalogue access tool and SIMBAD database operated at CDS, Strasbourg, France, were invaluable resources for this work. 
This publication makes use of data products from the Two Micron All Sky Survey, which is a joint project of the University of Massachusetts and the Infrared Processing and Analysis Center/California Institute of Technology, funded by the National Aeronautics and Space Administration and the National Science Foundation.

This work benefited from involvement in ExoExplorers, which is sponsored by the Exoplanets Program Analysis Group (ExoPAG) and NASA’s Exoplanet Exploration Program Office (ExEP).
Caprice Phillips thanks the LSSTC Data Science Fellowship Program, which is funded by LSSTC, NSF Cybertraining Grant $\#$1829740, the Brinson Foundation, and the Moore Foundation; her participation in the program has benefited this work.
This work has made use of data from the European Space Agency (ESA) mission
{\it Gaia} (\url{https://www.cosmos.esa.int/gaia}), processed by the {\it Gaia}
Data Processing and Analysis Consortium (DPAC,
\url{https://www.cosmos.esa.int/web/gaia/dpac/consortium}). Funding for the DPAC
has been provided by national institutions, in particular the institutions
participating in the {\it Gaia} Multilateral Agreement

\bibliography{sample631}{}
\bibliographystyle{aasjournal}

\figsetstart
\figsetnum{1}
\figsettitle{Transmission Spectroscopy of Targets in Study}
\figsetgrpstart
\figsetgrpnum{1.1}
\figsetgrptitle{Top: Simulated NIRSpec/G235M, NIRSpec/G395M, and NIRISS (SOSS) transmission spectrum of LP 791-18~c with 10 transits (MMW $\sim$ 4.46). Bottom: S/N and deviation from flat line.}
\figsetplot{sn_figure_lp791_18c.pdf}
\figsetgrpnote{Top: Simulated NIRSpec/G235M, NIRSpec/G395M, and NIRISS (SOSS) transmission spectrum of each target with 10 transits (MMW $\sim$ 4.46). Bottom: S/N and deviation from flat line. The complete figure set (6 images) is available in the online journal.}
\figsetgrpend

\figsetgrpstart
\figsetgrpnum{1.2}
\figsetgrptitle{	Top: Simulated NIRSpec/G235M, NIRSpec/G395M, and NIRISS (SOSS) transmission spectrum of GJ 143~b with 10 transits (MMW $\sim$ 4.46). Bottom: S/N and deviation from flat line.}
\figsetplot{sn_figure_gj_143b.pdf}
\figsetgrpnote{Top: Simulated NIRSpec/G235M, NIRSpec/G395M, and NIRISS (SOSS) transmission spectrum of each target with 10 transits (MMW $\sim$ 4.46). Bottom: S/N and deviation from flat line. The complete figure set (6 images) is available in the online journal.}
\figsetgrpend

\figsetgrpstart
\figsetgrpnum{1.3}
\figsetgrptitle{	Top: Simulated NIRSpec/G235M, NIRSpec/G395M, and NIRISS (SOSS) transmission spectrum of TOI-270 d with 10 transits (MMW $\sim$ 4.46). Bottom: S/N and deviation from flat line.}
\figsetplot{sn_figure_toi_270d.pdf}
\figsetgrpnote{Top: Simulated NIRSpec/G235M, NIRSpec/G395M, and NIRISS (SOSS) transmission spectrum of each target with 10 transits (MMW $\sim$ 4.46). Bottom: S/N and deviation from flat line. The complete figure set (6 images) is available in the online journal.}
\figsetgrpend

\figsetgrpstart
\figsetgrpnum{1.4}
\figsetgrptitle{	Top: Simulated NIRSpec/G235M, NIRSpec/G395M, and NIRISS (SOSS) transmission spectrum of K2-18~b with 10 transits (MMW $\sim$ 4.46). Bottom: S/N and deviation from flat line.}
\figsetplot{sn_figure_k2_18b.pdf}
\figsetgrpnote{Top: Simulated NIRSpec/G235M, NIRSpec/G395M, and NIRISS (SOSS) transmission spectrum of each target with 10 transits (MMW $\sim$ 4.46). Bottom: S/N and deviation from flat line. The complete figure set (6 images) is available in the online journal.}
\figsetgrpend

\figsetgrpstart
\figsetgrpnum{1.5}
\figsetgrptitle{	Top: Simulated NIRSpec/G235M, NIRSpec/G395M, and NIRISS (SOSS) transmission spectrum of K2-3~c with 10 transits (MMW $\sim$ 4.46). Bottom: S/N and deviation from flat line.}
\figsetplot{sn_figure_k2_3c.pdf}
\figsetgrpnote{Top: Simulated NIRSpec/G235M, NIRSpec/G395M, and NIRISS (SOSS) transmission spectrum of each target with 10 transits (MMW $\sim$ 4.46). Bottom: S/N and deviation from flat line. The complete figure set (6 images) is available in the online journal.}
\figsetgrpend

\figsetgrpstart
\figsetgrpnum{1.6}
\figsetgrptitle{	Top: Simulated NIRSpec/G235M, NIRSpec/G395M, and NIRISS (SOSS) transmission spectrum of LHS 1140~b with 10 transits (MMW $\sim$ 4.46). Bottom: S/N and deviation from flat line.}
\figsetplot{sn_figure_LHS_1140b.pdf}
\figsetgrpnote{Top: Simulated NIRSpec/G235M, NIRSpec/G395M, and NIRISS (SOSS) transmission spectrum of each target with 10 transits (MMW $\sim$ 4.46). Bottom: S/N and deviation from flat line. The complete figure set (6 images) is available in the online journal.}
\figsetgrpend

\figsetend

\begin{figure}
\figurenum{1}
\plotone{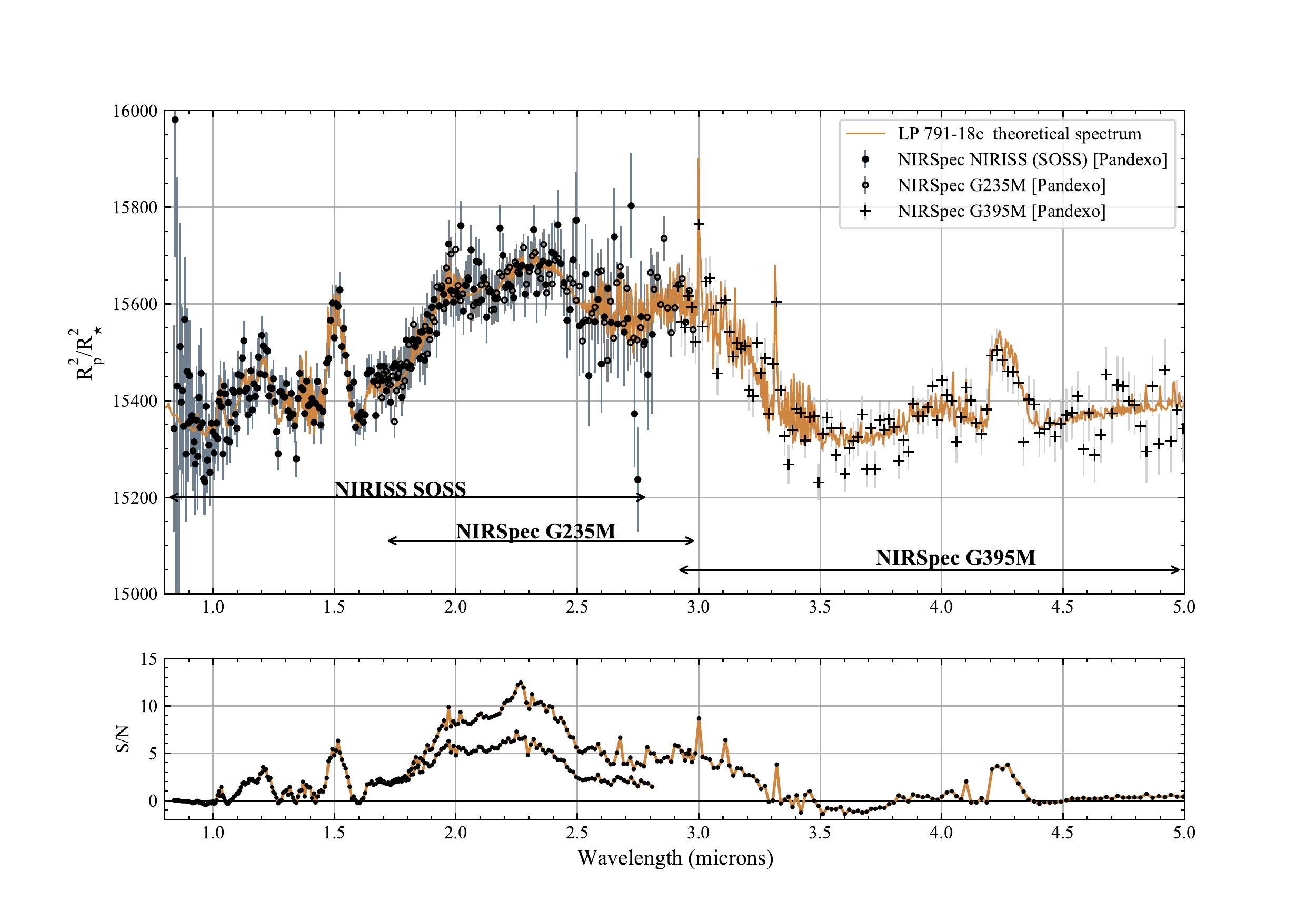}
\caption{Top: Simulated NIRSpec/G235M, NIRSpec/G395M, and NIRISS (SOSS) transmission spectrum of LP 791-18~c with 10 transits (MMW $\sim$ 4.46). Bottom: S/N and deviation from flat line. The complete figure set (6 images) is available in the online journal.}
\end{figure}

\end{CJK*}

\end{document}